\documentclass[
 aip,
amsmath,amssymb,
reprint,twocolumn 
]{revtex4-1}

\usepackage{graphicx}
\usepackage{dcolumn}
\usepackage{bm}

\usepackage[utf8]{inputenc}
\usepackage[T1]{fontenc}
\usepackage{mathptmx}
\usepackage{etoolbox}
\usepackage{xcolor}
\usepackage{amssymb}
\usepackage{amsfonts}
\usepackage{amsmath}
\usepackage{longtable}
\usepackage{epsfig}
\usepackage{float}

\def\dg{\dagger}
\newcommand{\circnum}[1]{\textcircled{#1}} 
\renewcommand{\circnum}[1]{{\bf #1.}}  

\makeatletter
\def\@email#1#2{%
	\endgroup
	\patchcmd{\titleblock@produce}
	{\frontmatter@RRAPformat}
	{\frontmatter@RRAPformat{\produce@RRAP{*#1\href{mailto:#2}{#2}}}\frontmatter@RRAPformat}
	{}{}
}%
\makeatother
\begin{document}
	
	\preprint{AIP/123-QED}
	
	\title{Influence of induced excitation on the functionality of finite--length molecular chain}
	\author{D. Chevizovich}
	\affiliation{Vinca Institute of Nuclear Sciences--National Institute of the Republic of Serbia, University of Belgrade, P.O. BOX 522, 11001, Belgrade, Serbia}
	\email{cevizd@vin.bg.ac.rs}
	\author{S. Galovic}
	\affiliation{Vinca Institute of Nuclear Sciences--National Institute of the Republic of Serbia, University of Belgrade, P.O. BOX 522, 11001, Belgrade, Serbia}
	\author{V. Matic}
	\affiliation{Vinca Institute of Nuclear Sciences--National Institute of the Republic of Serbia, University of Belgrade, P.O. BOX 522, 11001, Belgrade, Serbia}		
	\author{Z. Ivic}
	\affiliation{Vinca Institute of Nuclear Sciences--National Institute of the Republic of Serbia, University of Belgrade, P.O. BOX 522, 11001, Belgrade, Serbia}
	\affiliation{Institute of Theoretical and Computational Physics, Physics Department University of Crete, Voutes Campus GR--70013 Voutes, Heraklion Crete, Greece}
	\author{Z. Przulj}	
	\affiliation{Vinca Institute of Nuclear Sciences--National Institute of the Republic of Serbia, University of Belgrade, P.O. BOX 522, 11001, Belgrade, Serbia}
	\affiliation{Faculty of Electrical Engineering, University of East Sarajevo, Vuka Karadzica 30, East Sarajevo, Republic of Srpska, Bosnia and Herzegovina}%

	\date{\today}

	\maketitle

	\section{\label{Abstract} Abstract}

This study investigates the possibility that an injected intramolecular excitation, which becomes self--trapped due to its interaction with thermal oscillations of the molecular chain, can influence the chain's functionality in physiological processes in which it eventually participates. It is assumed that the presence of an excitation at a given node locally alter the physical properties of the chain, such as the electric dipole moment or charge distribution, thereby potentially disrupting its biochemical functions. Quantum resonance effects may cause the excitation, initially induced at one structural element, to delocalize and reappear at a distant site. To explore this phenomenon, we developed and analyzed a theoretical model in which a single excitation is injected in a specific structural element of a finite molecular chain in thermal equilibrium with its environment. Differential equations for the correlation functions were derived and solved analytically, providing the determination of the probability of finding the excitation at each site. The results indicate that injected excitation can affect the function of the biomolecule, especially at higher temperatures, when the excitation residence time on a molecular node becomes long enough to affect the functionality of the local segment of the biomolecule. Moreover, the position of the initially excited node can lead to the emergence of an asymmetric probability distribution with respect to the initial site, resulting in an effective "directed" migration of the excitation.	

	\section{\label{sec:level2} Introduction}
	
	Biomolecular chains (BmC) participate in a variety of biochemical processes within living cells. Notable examples include processes in which quanta of energy or charged particles are transferred along the chain (e.g., during cellular respiration). Other examples are processes in which a molecule functions as a carrier of information that is conveyed to another molecule (such as encoding, storage, and transfer of bioinformation, or molecular recognition processes); and processes in which a biomolecule or its part serves as a trigger or regulator for other biochemical reactions (enzymes, catalysts) \cite{Voet, DavydovBQM, Lehninger, Petrov, Dauxois, Frohlich, CruzeiroLTP, RSOS, ChenACIE}. The outcome of these processes depends not only on the intrinsic properties of the biomolecule, such as its composition or geometry, but also on the physical state of its local segments at the moment when one molecule encounters another. It is evident that these processes are highly efficient; otherwise, life would not be possible. For example, the transport of energy quanta in the form of intramolecular vibrational excitations of the C=O group of peptide bonds (vibrons) along polypeptide chains occurs with an efficiency exceeding 90\% (Fig. 1). Similarly high efficiencies are observed for the migration of excited electrons at the submolecular level during cellular respiration and photosynthesis.
		
	\begin{figure}[h]
		\begin{center}
			\includegraphics[width=80mm]{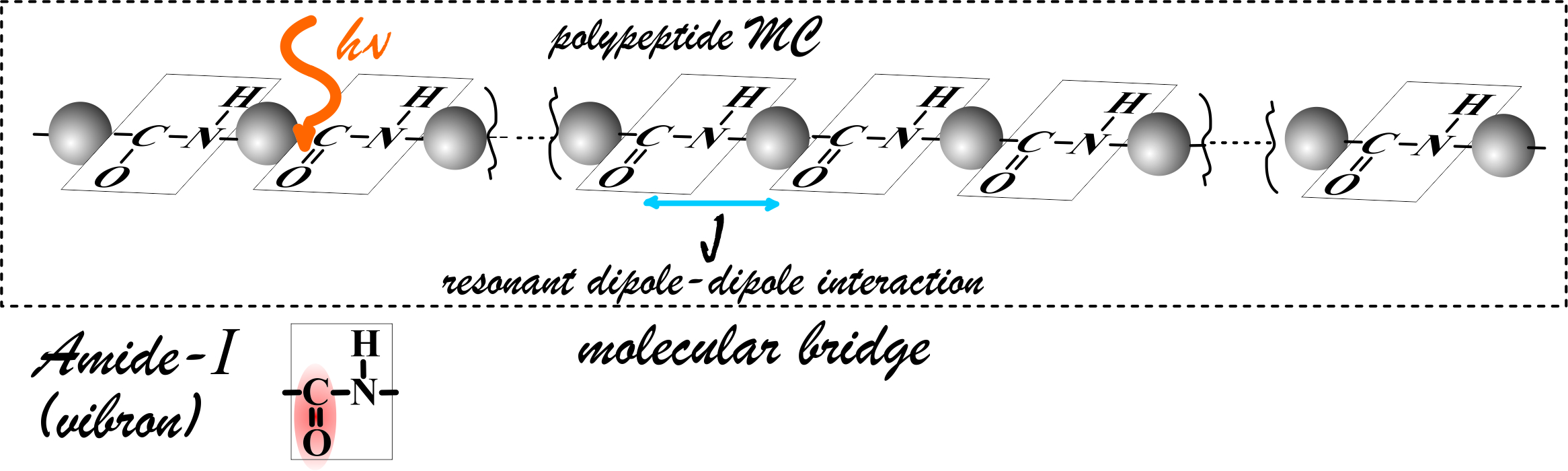}
			\vspace{-3mm}
			\caption{Understanding the excitation migration through BmC is particularly important in the bioenergetics of living cells. The figure shows a schematic representation of energy quantum transport along the polypeptide MC.}\label{fig01}
		\end{center}
	\end{figure}
		
	However, the appearance of an excitation at a structural element (SE) of a BmC may locally alter its properties, such as the distribution of charges, electric dipole moments, or even the local geometry of the molecule. If the excitation persists in the segment containing this structural element long enough and with sufficiently high probability, the functionality of that segment in the biochemical processes in which it participates may be impaired or even completely disrupted. Furthermore, an excitation initially induced at a particular SE of a molecular chain (MC) may delocalize and appear on a distant segment of the BmC \cite{DavydovBQM, Petrov, CruzeiroLTP, RSOS, ChenACIE, DavydovSMS, PetrovJTB, BrizhikPRE2014, HolsteinAP1, HolsteinAP2, ScottPR, Schuttler, CastroNetoCaldeira, ChevizovichPEPAN2025}. This effect is especially important in complex biomolecules such as proteins, DNA, and RNA, where different segments of the molecule are involved in distinct physiological functions \cite{Voet, Lehninger,Petrov, Dauxois, RSOS}. The success of excitation migration depends on the basic properties of the biomolecule and the physical state of its segments when the molecule carrying the excitation encounters another molecule.	
	
	
	One should bear in mind that an excitation, once it appears in a BmC, interacts with the thermal oscillations of its SEs, while the chain itself is assumed to be in thermal equilibrium with its environment \cite{HolsteinAP1, HolsteinAP2, LF, AK, AlvermanPRB, YarkonyJCP, ZdravkovicCevizovicND}. As a result of this interaction, the excitation may, in general, dissipate its energy (for example, into the vibrational modes of the BmC), or even completely lose it, causing the excitation itself to disappear. From the standpoint of classical physics, such a scenario is inevitable. Consequently, the stability of the states of an induced excitation in a BmC is a phenomenon that is difficult to explain within this framework. In fact, the mechanisms that ensure the stability of excitations in BmCs, as well as the high efficiency of processes in which their migration plays a key role, thereby contributing to the stability of physiological processes in living cells, have been the subject of scientific investigation for decades. Numerous physical models have been developed to address this issue \cite{DavydovBQM, BrizhikPRE2014, HolsteinAP1, HolsteinAP2, LF, AK, YarkonyJCP, ZdravkovicCevizovicND, CevizovicPRE, FalvoPouthier, EminPRB, GogolinPR1988, BI, HammTsironisEPJST147, HammTsironis, IvicCP426}. However, none of the existing models has so far succeeded in providing a comprehensive description applicable to different types of excitations, particularly to their transport across large intramolecular distances. It should be emphasized that the stability of such processes gave rise to the opinion that excitation migration at the submolecular level is a quantum--mechanical effect, and that its explanation (and modeling) should be based on the principles of quantum mechanics. Among the first to put forward this idea was Albert Szent--Gy{\"o}rgyi \cite{ASG}, as early as 1957. From the quantum mechanical point of view, besides causing energy loss, the interaction of an induced excitation with thermal oscillations of the BmC  can cause the excitation to become self--trapped, forming a polaron state-- a new quasiparticle, "dressed" by the cloud of virtual phonons \cite{DavydovBQM, AK, CevizovicPRE, HammTsironis, FalvoPouthier}. The remaining weak interaction with residual phonons allows the polaron to behave as a quasi--free quantum particle and to sustain quantum coherence. The formed ST excitation, due to resonant dipole--dipole interactions between neighboring SEs (in the case of a vibron excitation), or due to overlap of electronic orbitals of adjacent SEs (if the excitation involves an electron in a higher energy molecular orbitals), can form a delocalized state extending over the entire MC \cite{LF, AK, HolsteinAP1, HolsteinAP2, YarkonyJCP}. Thus, rather than dissipating energy through collisions with phonons, the excitation "migrates" almost freely along the BmC in the form of a dressed quasiparticle and may eventually localize on a SE located far from the site where it was initially created. Although both the ordinary excitation and the excitation in a polaron state can coexist \cite{ZdravkovicCevizovicND, CevizovicPRE, HammTsironisEPJST147}, the latter is capable of migrating over large intramolecular distances while preserving its quantum nature.
	
	In this work, we consider the possibility that an excitation, induced in a MC of finite length, can resonantly migrate through the structure and reach a distant SE, thereby affecting the function of the segment containing this remote SE. It is assumed that the MC is composed of identical SEs that interact with each other. We analyzed the probability of finding the excitation and its residence time on the distant SE of the MC. Moreover, it is assumed that the excitation, through its interaction with the thermal oscillations of the MC, forms a ST state corresponding to a small (non--adiabatic) polaron. The small polaron model is generally accepted as providing an adequate description of the vibron self--trapping process in polypeptide quasi--one--dimensional MCs \cite{LF, AK, YarkonyJCP, CevizovicPRE, HammTsironisEPJST147, HammTsironis, FalvoPouthier, PouthierJCP132, PouthierPRL, KalosakasPRB, PouthierPRE2008}. Such a ST excitation (an excitation "dressed" by a cloud of virtual phonons) exhibits properties that differ significantly from those of a "bare" excitation: its effective mass can be substantially larger than that of a "bare" excitation, while the residual interaction with renormalized phonons is considerably weaker and can be neglected \cite{LF, AK, YarkonyJCP, CevizovicPRE, FalvoPouthier, PouthierJCP132, PouthierPRL}.
	
	To determine and analyze the probability of finding the excitation on different SEs of the MC, in this work an analytical model was formulated. From this model, a system of differential equations for the correlation functions (CF) of the ST excitation on various structural elements of the MC was derived. The resulting system of coupled differential equations was solved using integral transform methods. In this way, the time--dependent probabilities of finding the excitation on distant SE of the MC were obtained, which depend on the temperature of the molecular chain's environment, its length (geometry), as well as on the values of the fundamental energy parameters of the system. The influence of the physical parameters characterizing the BmC, the environmental temperature, the chain length, and the position of the initially excited node of the BmC on the probability of the excitation appearance and residence time of the excitation on different SEs of the BmC was analyzed.

	\section{The model}
	
\begin{figure}[h]
		\begin{center}
			\includegraphics[width=80mm]{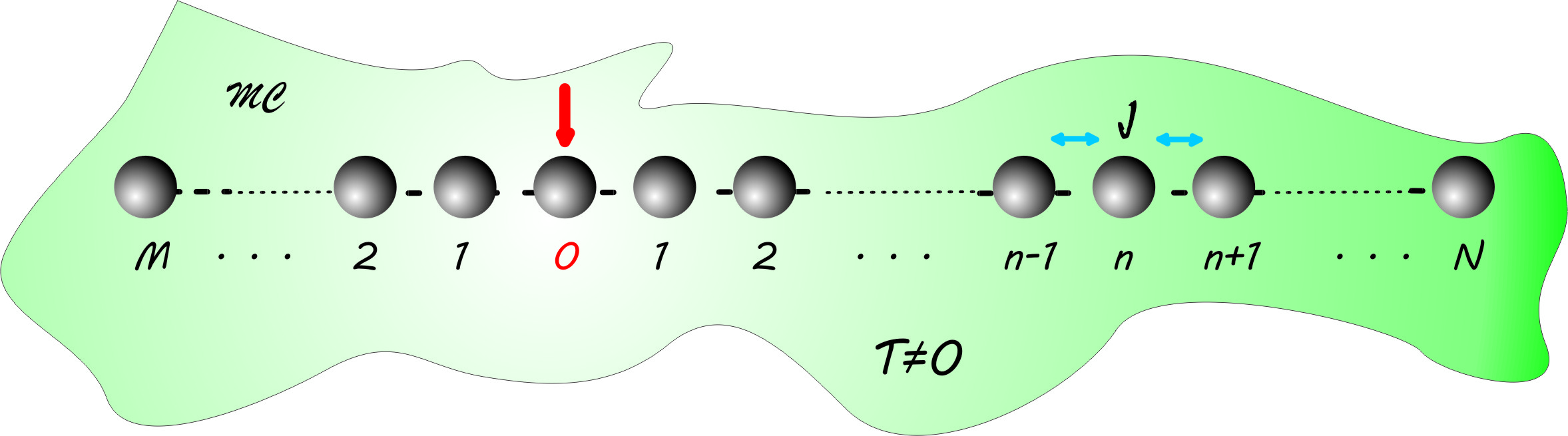}
			\caption{A schematic representation of the considered MC structure. The SE where the excitation is initially induced is labeled as $0$. On the left side of the $0$--th SE, there are $M$ SEs, numbered from 1 to $M$, while on the right side, there are $N$ SEs, numbered from 1 to $N$.}\label{fig02}
		\end{center}
	\end{figure}

	We considered a finite--size MC in thermal equilibrium with its surrounding medium (thermal bath). At the initial moment $t=0$, an excitation was induced at a particular SE of the MC, labeled as $0$ (the "zeroth" node of the MC) (see Fig.~\ref{fig02}). Let $N$ SEs be arranged to the right of the initially excited structural element, while $M$ SEs are arranged to its left. The SEs on the right are numbered consecutively from $1$ to $N$, starting from the "zeroth" node. Similarly, the SEs on the left side of the MC are numbered from $1$ to $M$, also starting from the "zeroth" node. Thus, the total number of SEs in the MC is $K=N+M+1$.
	
	The SE of the MC interact (and can exchange excitation) only with their nearest neighbors. We assumed that, due to the interaction with thermal oscillations of the SE, the excitation becomes self--trapped. Furthermore, due to the resonant interaction between neighboring SEs of the MC, the ST excitation may delocalize over the entire chain and, at a later time, appear at any location within the MC. To describe the migration of the ST excitation through the MC, we started from the Holstein model of the molecular crystal \cite{HolsteinAP1, HolsteinAP2, ZdravkovicCevizovicND,KalosakasPRB}, modified for the case of the finite--size MC.
	
	\begin{equation}\label{PocHam}
		\hat{H}=\hat{H}_{\text{exc}}+\hat{H}_{\text{ph}}+\hat{H}_{\text{exc--ph}}
	\end{equation}
	
	\noindent where
	
	\begin{equation}\label{PocHamExc}
	\hat{H}_{\text{exc}}=\mathcal{E}_0\sum_n\hat{a}^{\dag}_n\hat{a}_n-J_0\sum_n\hat{a}^{\dag}_n\left(\hat{a}_{n+1}+\hat{a}_{n-1}\right)
	\end{equation}
	
	\begin{equation}\label{PocHamPh}
	\hat{H}_{\text{ph}}=\sum_q\hbar\omega_q\hat{b}^{\dag}_q\hat{b}_q
	\end{equation}
		
	\begin{equation}\label{PocHamExcPh}
	\hat{H}_{\text{exc--ph}}=\frac{1}{\sqrt{N}}\sum_n\sum_qF_q\mathrm{e}^{iqnR_0}\hat{a}^{\dag}_n\hat{a}_n(\hat{b}_q+\hat{b}_{-q})
	\end{equation}
	
	\noindent Here, $\hat{a}^{\dg}_n$ ($\hat{a}_n$) are the creation (annihilation) operators of the excitation on the $n$--th SE of the MC, while $\hat{b}^{\dg}_q$ ($\hat{b}_q$) are the creation (annihilation) operators of phonons with wave number $q$. The energy required to excite the corresponding excitation mode on the particular SE of the MC is $\mathcal{E}_0$. The transfer integral (or the energy of the resonant dipole--dipole interaction in the vibron case) between adjacent SEs of the MC is denoted by $J$. Finally, excitation--phonon coupling function is $F_q=\chi\sqrt{\frac{\hbar}{2M\omega_0}}$ for optical phonon modes and $F_q=2i\chi\sqrt{\frac{\hbar}{2M\omega_q}}\sin(qR_0)$ for acoustic phonon modes ($\omega_q=\omega_0\sin{qR_0/2}$), where $qR_0\in\left[-\pi,\pi\right]$. The parameter $\chi$ is the excitation--phonon coupling parameter, which characterizes the strength of the interaction between the excitation and phonon modes  \cite{LF, AK, YarkonyJCP, CevizovicPRE, ZdravkovicCevizovicND, PouthierJCP132, PouthierPRL, KalosakasPRB, PouthierPRE2008, HennigPRB}. The characteristic phonon frequency is $\omega_0=2\sqrt{\frac{\kappa}{M}}$.
	
	As previously noted, we assumed that, due to the interaction with thermal oscillations of the SEs of the MC, the excitation forms a self--trapped state. If this state is sufficiently stable (meaning that the remaining ST excitation--phonon interaction is negligible), the ST excitation can migrate along the molecular chain practically without energy loss \cite{DavydovBQM, AK, LF, YarkonyJCP, CevizovicPRE, ZdravkovicCevizovicND, PouthierJCP132, ZdravkovicCevizovicND, Rashba}. The transition to the picture of ST excitation (or, more generally, to the polaron picture) was achieved via the Lang--Firsov unitary transformation \cite{LF, ZdravkovicCevizovicND, CevizovicPRE, PhysB2005Ivic, BI, Mahan} (LFuT): $$\hat{U}_{\text{LF}}=\mathrm{e}^{-\sum_n\hat{a}^{\dag}_n\hat{a}_n\hat{S}_n},\quad \hat{S}_n=\frac{1}{\sqrt{N}}\sum_q \frac{F^*_q}{\hbar\omega_q} \mathrm{e}^{-iqnR_0}\left(\hat{b}_{-q}-\hat{b}_q\right)$$ Transformed Hamiltonian $\hat{H}_{\text{LF}}=\hat{U}_{\text{LF}}\hat{H}\hat{U}^{\dag}_{\text{LF}}$ is
	
 	\begin{align}\label{HLF}
	\hat{H}_{\text{LF}}&=\mathcal{E}_R\sum_n\hat{a}^{\dag}_n\hat{a}_n-J_0\sum_n\hat{a}^{\dag}_n\left(\hat{a}_{n+1}\hat{T}_+(n)+\hat{a}_{n-1}\hat{T}_-(n)\right)+\nonumber\\
	&+\sum_q\hbar\omega_q\hat{b}^{\dag}_q\hat{b}_q+\hat{O}_{\text{R--R}}\nonumber
	\end{align}
	
	\noindent Here, $\hat{a}^{\dg}_n$ ($\hat{a}_n$) are the creation (annihilation) operators of the ST excitation on the $n$--th SE, and $\hat{b}^{\dg}_q$ ($\hat{b}_q$) are those of the new (renormalised) phonons. The energy of the ST excitation is $\mathcal{E}_R=\mathcal{E}_0-\mathcal{E}_b$, with excitation binding energy \cite{LF,YarkonyJCP,CevizovicPRE,ZdravkovicCevizovicND} $\mathcal{E}_b=-\frac{1}{N}\sum_q\frac{|F_q|^2}{\hbar\omega_q}$. The dressing mechanism is incorporated via the application of the LFuT, which leads to appearance of the nonlinear operators $\hat{T}_{\pm}(n)=\mathrm{e}^{\hat{S}_{n\pm 1}-\hat{S}_n}$. The LFuT provides an exact diagonalization of the Hamiltonian in the limit $J_0\to 0$. However, when $J_0 \neq 0$, a nonlinear coupling  $\hat{a}^{\dag}_n\hat{a}_{n\pm 1}\hat{T}_{\pm}(n)$ remains, leading to dissipation processes and incoherent motion of the ST excitation. The term $\hat{O}_{\text{R--R}}$ represents "residual" interaction between two ST excitations is neglected in the single excitation case.
	
	Assuming thermal equilibrium, the ambient temperature determines the mechanical oscillations of the SEs, which indirectly affect the state of the ST excitation via phonon--excitation coupling. To account for the influence of thermal fluctuations on the properties of the ST excitation, we applied the mean--field approximation and find the mean--field Hamiltonian of our system\cite{LF, AK, YarkonyJCP, CevizovicPRE, ZdravkovicCevizovicND}: $$\hat{H}_{\text{MF}}=\hat{\mathcal{H}}_{R}+\hat{\mathcal{H}}_{\text{ph}}+\hat{\mathcal{H}}_{\text{rest}}$$ with the ST excitation Hamiltonian in site representation:
	
	\begin{equation}\label{HSP}
	\hat{\mathcal{H}}_R=\mathcal{E}_R\sum_n\hat{a}^{\dag}_n\hat{a}_n-J_0\mathrm{e}^{-W(T)}\sum_n\hat{a}^{\dag}_n\left(\hat{a}_{n+1}+\hat{a}_{n-1}\right)
	\end{equation}
	
	\noindent The term $\hat{\mathcal{H}}_{\text{rest}}$ includes residual (weak) ST excitation--phonon and ST excitation--ST excitation interactions, which are neglected. The thermal average of the nonlinear operators $\hat{T}_{\pm}(n)$ is given by $\left\langle\hat{T}_{\pm}(n)\right\rangle_{\text{ph}} =\mathrm{e}^{-W(T)}$, where the narrowing factor of $J_0$ is $$W(T)=\frac{1}{N}\sum_q\frac{|F_q|^2}{(\hbar\omega_q)^2} (2n_q+1)(1-\cos(qR_0)),\;\; n_q=\frac{1}{\mathrm{e}^{\frac{\hbar\omega_q}{k_BT}}-1}.$$ Here, $\left\langle\hat{A}\right\rangle_{\text{ph}}$ denotes the average of operator $\hat{A}$ over the renormalized phonon ensemble, which is in thermal equilibrium with a thermal bath at temperature $T$. Another important effect of excitation self--trapping is the reduction of the transfer integral $J_0$. Although its value is reduced, the fact that it remains nonzero allows the excitation to migrate along the MC, making it possible for the excitation to appear on any SE of the MC. The Hamiltonian in Eq.(\ref{HSP}) provides the mathematical basis for further investigation of excitation migration. In the following text, we focus on the vibron excitation that interacts with optical phonon modes. Specifically, it is believed that in the case of ST of the Amide--I excitation in polypeptide molecular structures, the interaction with optical phonons plays a crucial role in the formation of the polaron state \cite{AK}.

	\section{Probability of the excitation appearing on nodes of the MC}
	
	To analyze the appearance of the excitation on different SEs of the MC, let us assume that at the initial moment $t=0$ the excitation is induced on a particular SE labeled by the index $0$. The initial excitation state is $\left|\psi_i(0)\right\rangle= \hat{a}_0^{\dg}\left|0\right\rangle$, where $\left|0\right\rangle$ represents the excitation "vacuum" state. In the absence of external influences, this state evolves in time as $\left|\psi_i(t)\right\rangle= \mathrm{e}^{-\frac{i}{\hbar}t\hat{\mathcal{H}}_R}\hat{a}_0^{\dg}\left|0\right\rangle$. Now, we are interested in the probability of appearing the excitation on a distant node $n\neq 0$, at the latter time $t\neq 0$, i.e., the probability that the system reaches the state $\left|\psi_f(t)\right\rangle=\hat{a}^{\dg}_n\left|0(t)\right\rangle$. Here, $\left|0(t)\right\rangle=\mathrm{e}^{-\frac{i}{\hbar}t\hat{\mathcal{H}}_R}\left|0\right\rangle$ describes the time evolution of the excitation vacuum state. This probability is given by $p_n(t)=|\left\langle\psi_f(t)|\psi_i(t)\right\rangle|^2$, with the corresponding correlation function (CF) given by $V_n(t)=\left\langle\psi_f(t)|\psi_i(t)\right\rangle=\left\langle 0\right|\mathrm{e}^{\frac{i}{\hbar}t\hat{\mathcal{H}}_R}\hat{a}_n\mathrm{e}^{-\frac{i}{\hbar}t\hat{\mathcal{H}}_R}\hat{a}_0^{\dg}\left|0\right\rangle$. In the Heisenberg picture it is given by:
	
	\begin{equation}\label{CFVH}
	V_n(t)=\left\langle 0\right|\hat{a}_n(t)\hat{a}_0^{\dg}\left|0\right\rangle;\quad \hat{a}_0^{\dg}=\hat{a}_0^{\dg}(0)
	\end{equation}
	
	The CF in Eq.(\ref{CFVH}) determines the probability amplitude for the transition of an excitation from the molecular node where it was initially created (the "zeroth" node) to a distant $n$--th node, over the time interval from $t_0=0$ to a later time $t$. Mathematically, it links the creation of the excitation at the initial node at $t=0$ with its appearance at node $n$, at time $t$. For brevity, we will henceforth refer to the CF in Eq.(\ref{CFVH}) simply as the correlation function associated with node $n$ at time $t$. The following procedure consists of solving the system of differential equations for the CFs $V_n(t)$ for all nodes of the MC.

	\subsection{Differential equations for the correlation functions $V_n(t)$}
	
	To derive the differential equations for the CF, we consider a finite--length MC consisting of $K=N+M+1$ SEs, as described at the beginning of Section III and illustrated in Fig.~\ref{fig02}. The CFs associated with the nodes on the right side of the "zeroth" node are labeled as $V_n(t)$ and indexed by $n \in {1,2,\ldots,N}$. The CFs associated with the nodes on the left side of the "zeroth" node are labeled as $U_m(t)$ and indexed by $m \in {1,2,\ldots,M}$. For the CF associated with the "zeroth" node, we adopt the convention $U_0 = V_0$. Furthermore, we introduce the dimensionless "reduced" time variable via the substitution $\tau=\omega_0t$. In this way, all time derivatives transform as $\frac{df(t)}{dt}=\omega_0\frac{dF(\tau)}{d\tau}$. Finally, we introduce a special notation for the reduced transfer integral, $J = J_0 \mathrm{e}^{-W(T)}$. With these definitions, the system of differential equations for the CFs is readily obtained:
	
	\begin{equation}\label{dV0}
		i\frac{d{V}_0}{d\tau}=\frac{\mathcal{E}_R}{\hbar\omega_0}V_0-\frac{J}{\hbar\omega_0}(U_1+V_1)
	\end{equation}
	
	\noindent for the CF associated with "zeroth" node of the MC,
	
	\begin{align}\label{dVdesno}
	i\frac{d{V}_1}{d\tau}&=\frac{\mathcal{E}_R}{\hbar\omega_0}V_1-\frac{J}{\hbar\omega_0}(V_0+V_2)\nonumber\\
	..............&............................................\nonumber\\
	i\frac{d{V}_{N-1}}{d\tau}&=\frac{\mathcal{E}_R}{\hbar\omega_0}V_{N-1}-\frac{J}{\hbar\omega_0}(V_{N-2}+V_N)\\
	i\frac{d{V}_N}{d\tau}&=\frac{\mathcal{E}_R}{\hbar\omega_0}V_N-\frac{J}{\hbar\omega_0}V_{N-1}\nonumber
	\end{align}
	
	\noindent for the CF associated with nodes on the right side of the "zeroth" node,
	
	\begin{align}\label{dVlijevo}
	i\frac{d{U}_1}{d\tau}&=\frac{\mathcal{E}_R}{\hbar\omega_0}U_1-\frac{J}{\hbar\omega_0}(U_2+V_0)\nonumber\\
	..............&............................................\nonumber\\
	i\frac{d{U}_{M-1}}{d\tau}&=\frac{\mathcal{E}_R}{\hbar\omega_0}U_{M-1}-\frac{J}{\hbar\omega_0}(U_{M-2}+U_M)\\
	i\frac{d{U}_M}{d\tau}&=\frac{\mathcal{E}_R}{\hbar\omega_0}U_M-\frac{J}{\hbar\omega_0}U_{M-1} \nonumber
	\end{align}
	
	\noindent for the CF associated with nodes on the left side of the "zeroth" node. At the initial moment, the excitation is induced at the "zeroth" node:
	
	\begin{equation}\label{IntCond}
		|V_0(\tau=0)|^2=1
	\end{equation}
		
	\noindent For the sake of clarity and readability, the time dependence of the CFs $V_i(\tau)$ and $U_i(\tau)$ has been omitted in the above equations. Wherever this dependence does not appear explicitly, it is implicitly assumed. The obtained system of equations represents a set of coupled, first--order differential equations in time. It can be solved analytically using the method of integral transformations. More precisely, to solve them we applied the Laplace transformation (LT) method: $$\mathcal{L}\left\{f(\tau)\right\}=F(s)= \int_0^{+\infty}f(\tau)\mathrm{e}^{-s\tau}d\tau;\quad \tau\ge 0$$ where $s\in\mathbb{C}$ is complex frequency. By applying the LT to the system of equations Eq.(\ref{dV0}), Eq.(\ref{dVdesno}), and Eq.(\ref{dVlijevo}), we obtain a system of algebraic equations for the LT of the CFs:
	
	\begin{equation}\label{Vlin0}
	is\tilde{V}_0=\frac{\mathcal{E}_R}{\hbar\omega_0}\tilde{V}_0-\frac{J}{\hbar\omega_0}(\tilde{U}_1+\tilde{V}_1)+iV_0(\tau=0)
	\end{equation}
	
	\begin{align}\label{dVlindesno}
	is\tilde{V}_1&=\frac{\mathcal{E}_R}{\hbar\omega_0}\tilde{V}_1-\frac{J}{\hbar\omega_0}(\tilde{V}_0+\tilde{V}_2)\nonumber\\
	...........&......................................\nonumber\\
	is\tilde{V}_{N-1}&=\frac{\mathcal{E}_R}{\hbar\omega_0}\tilde{V}_{N-1}-\frac{J}{\hbar\omega_0}(\tilde{V}_{N-2}+\tilde{V}_N)\\
	is\tilde{V}_N&=\frac{\mathcal{E}_R}{\hbar\omega_0}\tilde{V}_N-\frac{J}{\hbar\omega_0}\tilde{V}_{N-1}\nonumber
	\end{align}
	
	\begin{align}\label{dVlinlijevo}
	is\tilde{U}_1&=\frac{\mathcal{E}_R}{\hbar\omega_0}\tilde{U}_1-\frac{J}{\hbar\omega_0}(\tilde{U}_2+\tilde{V}_0)\nonumber\\
	...........&......................................\nonumber\\
	is\tilde{U}_{M-1}&=\frac{\mathcal{E}_R}{\hbar\omega_0}\tilde{U}_{M-1}-\frac{J}{\hbar\omega_0}(\tilde{U}_{M-2}+\tilde{U}_M)\\
	is\tilde{U}_M&=\frac{\mathcal{E}_R}{\hbar\omega_0}\tilde{U}_M-\frac{J}{\hbar\omega_0}\tilde{U}_{M-1}\nonumber
	\end{align}
	
	\noindent For the sake of clarity and readability, the dependence of the LT of the CFs $\tilde{V}_i(s)$ and $\tilde{U}_i(s)$ on $s$ has been omitted in the above equations. Wherever this dependence does not appear explicitly, it is implicitly assumed. Introducing the complex variable
	
	\begin{equation}\label{x}
		x=\frac{\hbar\omega_0}{J} \left(-is+\frac{\mathcal{E}_R}{\hbar\omega_0}\right)
	\end{equation}
	
	\noindent and applying the expressions Eq.(\ref{V0}) and Eq.(\ref{bk2}) from Appendix I and Appendix II, the system of equations is solved and expressed in terms of \textit{modified Chebyshev polynomials of the second kind}\cite{AbramovicStegun}, $D_n(x)$:

	\begin{equation}\label{Vx}
		\tilde{V}_n(x)=-i\frac{\hbar\omega_0}{J}\frac{D_{N-n}(x)D_M(x)}{D_{M+N+1}(x)}V(\tau=0)
	\end{equation}
	
	\noindent $\forall n\in\left\{0,1,2,...,N\right\}$. We can now calculate the CF and, subsequently, the probabilities of finding the excitation at each individual SE of the MC as functions of time, the geometry of the MC, and the physical parameters of the system. For that purpose, the expression Eq.(\ref{Vx}) must be transformed into the time domain by applying the inverse LT. This requires changing from the variable $x$ back to the variable $s$ according to Eq.(\ref{x}). In addition, one needs to determine the singular points of the functions in Eq.(\ref{Vx}), that is, to find the roots of the polynomials appearing in their denominators. Once these roots are identified, the polynomials can be factorized as: $D_{M+N+1}(x)=(x-x_1)(x-x_2)\ldots(x-x_j)\ldots$ where $x_j$ are the roots of the polynomial $D_{M+N+1}(x)$. In the general case, we can remark that:
	
	\begin{equation}\label{fxTOs}
	(x-x_j)\rightarrow -i\frac{\hbar\omega_0}{J}\left(s+i\frac{J}{\hbar\omega_0}\left[\frac{\mathcal{E}_R}{\bar{J}}-x_j\right]\right)
	\end{equation}
	
	\noindent so that the factorization can be rewritten in the form $D_{M+N+1}(s)=(s-i\beta_1)(s-i\beta_2)\cdot...\cdot(s-i\beta_j)\cdot ...$. Here, $\beta_j$ are determined by Eq.(\ref{fxTOs}) and take the form: 
	
	\begin{equation}\label{betaj}
	\beta_j=\frac{J}{\hbar\omega_0}\left(x_j-\frac{\mathcal{E}_R}{J}\right),\; \beta_j\in\mathbb{R}
	\end{equation}	
	
	\noindent In the next step, the expression in Eq.(\ref{Vx}) is decomposed into a sum $$\frac{D_{N-n}(s)D_M(s)}{D_{M+N+1}(s)}=\frac{A_1}{s-i\beta_1}+\frac{A_2}{s-i\beta_2}+\frac{A_3}{s-i\beta_3}+...$$ where the coefficients $A_1$, $A_2$,... must be determined. This allows for a straightforward calculation of the inverse Laplace transform of the CF. Replacing the integration along the line $\Re{(s)}=\gamma$ with a suitable closed contour $C$, and applying the residue theorem:
	
	\begin{widetext}
	$$f(t)=\mathcal{L}^{-1}\left\{\frac{1}{s-i\beta}\right\}=\frac{1}{2\pi i} \int_{\gamma-i\infty}^{\gamma+i\infty}\mathrm{e}^{st}\frac{1}{s-i\beta}ds=\frac{1}{2\pi i}\oint_C\mathrm{e}^{st}\frac{1}{s-i\beta}ds=2\pi i\frac{1}{2\pi i} \operatorname{Res}_{s=i\beta}\left(\frac{\mathrm{e}^{st}}{s-i\beta}\right)=\mathrm{e}^{i\beta t}$$
	\end{widetext}

	\subsection{Probability of excitation appearance on distant SE}
	
	The form of the polynomials $\tilde{V}_n(x)$ depend on the MC length and the location at which the excitation is initially created. Consequently, to illustrate our model and analyze the probability distribution of the excitation appearance at distant nodes, it is convenient to consider a MC of finite length with the excitation initially localized at a specific site. As an example, we consider a MC comprising $K=11$ SEs, where the excitation at the initial time $t_0=0$ s is created on the second SE from the left. To the right of this element, there are $N=9$ SEs, while to the left there is $M=1$ element (Fig.\ref{fig02}). To investigate the probability of the excitation appearing on a distant SE, we examine the time evolution of the probability of finding the excitation on the rightmost SE: $p_9(t)=|V_9(t)|^2$. Although the calculation is demonstrated for a particular structural element, the probabilities of finding the excitation on other elements of the same molecular chain, or for different chain configurations (e.g., chains of different length or with the initial excitation at a different site), can be readily obtained using the procedure described below. We begin from the LT of the CF $V_9(x)$, defined in Eq.(\ref{Vx}):
	
	\begin{equation}\label{barV9x}
	\tilde{V}_9(x)=-i\frac{\hbar\omega_0}{J}\frac{D_0(x)D_1(x)}{D_{11}(x)}V(\tau=0)
	\end{equation}
	
	\noindent From Table \ref{tab:table2c}, we obtain the required polynomials $D_0(x)$, $D_1(x)$, and $D_{11}(x)$. By substituting them into Eq.~(\ref{barV9x}), we find
	
	\begin{equation}\label{barV9x2}
		\tilde{V}_9(x)=-i\frac{\hbar\omega_0}{J}\frac{x}{x^{11}-10x^9+36x^7-56x^5+35x^3-6x}V(\tau=0)\nonumber
	\end{equation}
	
	To find the CF $V_9(\tau)$, its Laplace transform $\tilde{V}_9(x)$ must be inverted back to the time domain by applying the inverse LT. In the first step, the denominator polynomial in Eq.(\ref{barV9x2}) is factorized into the form $(x-x_1)(x-x_2)\cdots(x-x_{11})$, where $x_j$ are the roots of the polynomial $D_{11}(x)$. Using Eq.(\ref{fxTOs}), the factorized denominator is then rewritten as $(s-i\beta_1)(s-i\beta_2)\cdots(s-i\beta_{11})$, where $\beta_j$ are given by Eq.(\ref{betaj}). Thus,
	
	\begin{equation}\label{V9s}
	\tilde{V}_9(s)=i\left(\frac{J}{\hbar\omega_0}\right)^9\frac{s+i\alpha}{(s-i\beta_1)(s-i\beta_2)...(s-i\beta_{11})}V(t=0)
	\end{equation}
	
	\noindent where $\alpha=\frac{\mathcal{E}_R}{\hbar\omega_0}$. Note that if the first root of $D_{11}(x)$ is chosen as $x_1=0$, then $\alpha=-\beta_1$. In the next step, the rational fraction in Eq.(\ref{V9s}) is decomposed into the sum $$\frac{s+i\alpha}{(s-i\beta_1)(s-i\beta_2)...(s-i\beta_{11})}=\frac{A_1}{s-i\beta_1}+\frac{A_2}{s-i\beta_2}+...+\frac{A_{11}}{s-i\beta_{11}}$$ and by multiplying both sides by $(s-i\beta_1)(s-i\beta_2)...(s-i\beta_{11})$ and successively setting $s=i\beta_1,\; i\beta_2,\; \ldots i\beta_{11}$, the expansion coefficients $A_j$ are readily obtained. Finally, we have:
	
	\begin{align}\label{Bj}
	A_1&=0\nonumber\\
	A_j&=-i\prod_{\substack{k\neq j}} \frac{1}{\beta_j-\beta_k}=-i\left(\frac{\hbar\omega_0}{J}\right)^9\underbrace{\prod_{\substack{k\neq j}}\frac{1}{x_j-x_k}}_{\mathbb{A}_j}
	\end{align} 
		
	\noindent where the indices $k, j$ take values in the set $\{2,3,...,11\}$, with $x_1$ taken as 0. Applying the inverse LT then yields
	
	\begin{equation}\label{V9t}
	V_9(\tau)=V(\tau=0)\sum_{j}\mathbb{A}_j\mathrm{e}^{i\beta_j\tau};\quad \mathbb{A}_j=\prod_{\substack{k\neq j}}\frac{1}{x_j-x_k}
	\end{equation}
	
	\noindent where $k,j\in\{2,3,...,11\}$. These results are most naturally expressed in terms of the dimensionless parameters $S=\frac{\mathcal{E}_b}{\hbar\omega_0}$ (coupling constant) and $B=\frac{2J_0}{\hbar\omega_0}$ (adiabatic parameter). With $\mathcal{E}_0=0$ (taken as the reference energy level) Eq.(\ref{betaj}) for $\beta_j$ becomes
	
	\begin{equation}\label{betajSB}
	\beta_j=S+\frac{B}{2}\mathrm{e}^{-W(\theta)}x_j 
	\end{equation}
	
	\noindent where
	
	\begin{equation}\label{Wtheta}
		W(\theta)=S\coth\left(1/2\theta\right)
	\end{equation}
	
	\noindent is the narrowing factor \cite{LF, YarkonyJCP, CevizovicPRE, ZdravkovicCevizovicND}, with $\theta=\frac{k_BT}{\hbar\omega_0}$ denoting the normalized temperature. Eq.(\ref{V9t}) then reads:
	
	\begin{equation}\label{V9tSB}
	V_9(\tau)=V(\tau=0)\mathrm{e}^{iS\tau}\sum_{j=2}^{11}\mathbb{A}_j\mathrm{e}^{i\tau\frac{B}{2}x_j\mathrm{e}^{-W(\theta)}}\nonumber
	\end{equation}
	
	Since the roots of $D_{11}(x)$ are symmetrically distributed around the root $x_1=0$,  we have adopted following numeration for remaining roots: $x_{2k+1}=-x_{2k}$ for $k\in\{1,2,...,5\}$. Because for $V_9(\tau)$ it holds that $\mathbb{A}_{2k+1}=-\mathbb{A}_{2k}$, we obtain: 	
	
	\begin{equation}\label{V9tSB2}
	V_9(\tau)=2iV(\tau=0)\mathrm{e}^{iS\tau}\sum_{k=1}^5{\mathbb{A}_{2k}\sin\left(\Omega_{2k}(\theta)\tau\right)}
	\end{equation}
	
	\noindent where 
	
	\begin{equation}\label{Omega}
	\Omega_i(\theta)=\frac{B}{2}x_i\mathrm{e}^{-W(\theta)},\quad i\in\{2,4,6,8,10\}
	\end{equation}
	
	Let us pay attention that CF $V_9(\tau)$ have the form of the sum of harmonics, where the harmonic frequencies depend on system parameters $B$ and $S$, the roots of the polynomial $D_i(x)$ as well as the values of the system temperature $\theta$. At the same time, the amplitude of each harmonic depends on the roots of the polynomial $D_i(x)$. The dependence of the amplitude on the parameter $S$ is "fictive", because it disappears when one calculate the probability of the excitation finding on the particular SE.

	\subsubsection{Analytical results: first observations}
	
	We can now make our first observations concerning the correlation function $V_9(\tau)$, i.e., the occurrence of the excitation at the last node of the MC. From Eqs.(\ref{V9tSB}) and (\ref{Omega}), one can conclude:\\
	
	\noindent\circnum{1} The correlation function $V_n(\tau)$, determining the probability distribution $v_n(\tau)$ can be represented as a sum of harmonics. Each harmonic corresponds to a positive root of the associated \textit{modified Chebyshev polynomial}. The number of harmonics contributing to the composition of the correlation function thus depends on the number of structural elements between the excited node and the furthest node, since the degree of this Chebyshev polynomial depends on $N$.\\
		
	\noindent\circnum{2} The frequency of each harmonic $\Omega_i$ is determined by the roots of the modified Chebyshev polynomials $x_i$ (i.e., by the MC geometry), but also depends on the exciton--phonon interaction strength (i.e., on the coupling constant $S$), the adiabatic parameter $B$ (i.e., transfer--integral $J_0$), and the system temperature $\theta$. Increasing $S$ or $\theta$ lowers the frequency and broadens the probability maximum for finding the excitation on the MC node, thereby prolonging exciton residence time on the node, whereas increasing $B$ raises the frequency, narrows the maximum, and shortens its residence time.\\	
	
	\noindent\circnum{3} The intensity of each harmonic is determined solely by the roots of the modified Chebyshev polynomials, meaning that $S$, $B$ and temperature do not affect the peak values of the probability distributions, only their width and position. In practice, the magnitude of the probability maximum for finding the excitation on a given MC node is set by the chain geometry (specifically, the relative positions of the zeroth and target nodes, as well as by the total length of the MC).\\
	
	To provide a more detailed illustration of the general conclusions from the analytical expressions Eq.(\ref{V9tSB2}) and Eq.(\ref{Omega}) and a clearer analysis of the model results, the following section presents the time dependence of the probability for finding the excitation on the last MC node, first for various system temperatures $\theta$ and then for different values of the coupling constant $S$.

	\subsection{Probability of excitation appearance on the most distant node of the MC}
	
	Let us now examine the graphs showing the probability of finding the excitation on the last node of a MC consisting of $K=11$ nodes, where the excitation is initially induced at the second SE of the chain, observed from the left end. We consider the time probability distributions for finding the excitation on the last node at the right end of the chain, as functions of the environmental temperature $\theta$ and the coupling strength $S$ between the dressed excitation and the thermal oscillations of the chain's SEs.

	\subsubsection{The influence of temperature $\theta$}
	
\begin{figure}[h]
		\begin{center}
			\includegraphics[width=4cm]{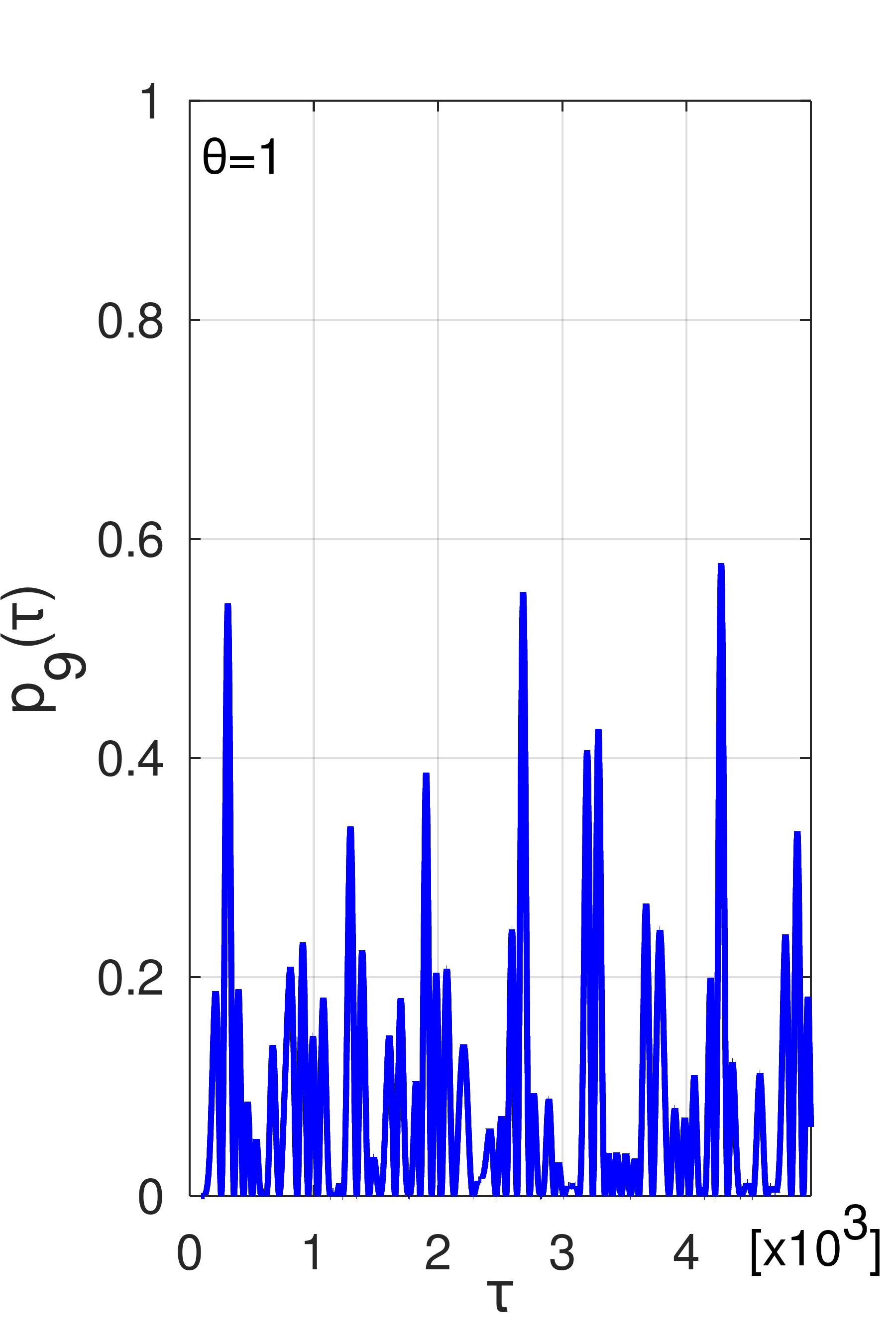}
			\includegraphics[width=4cm]{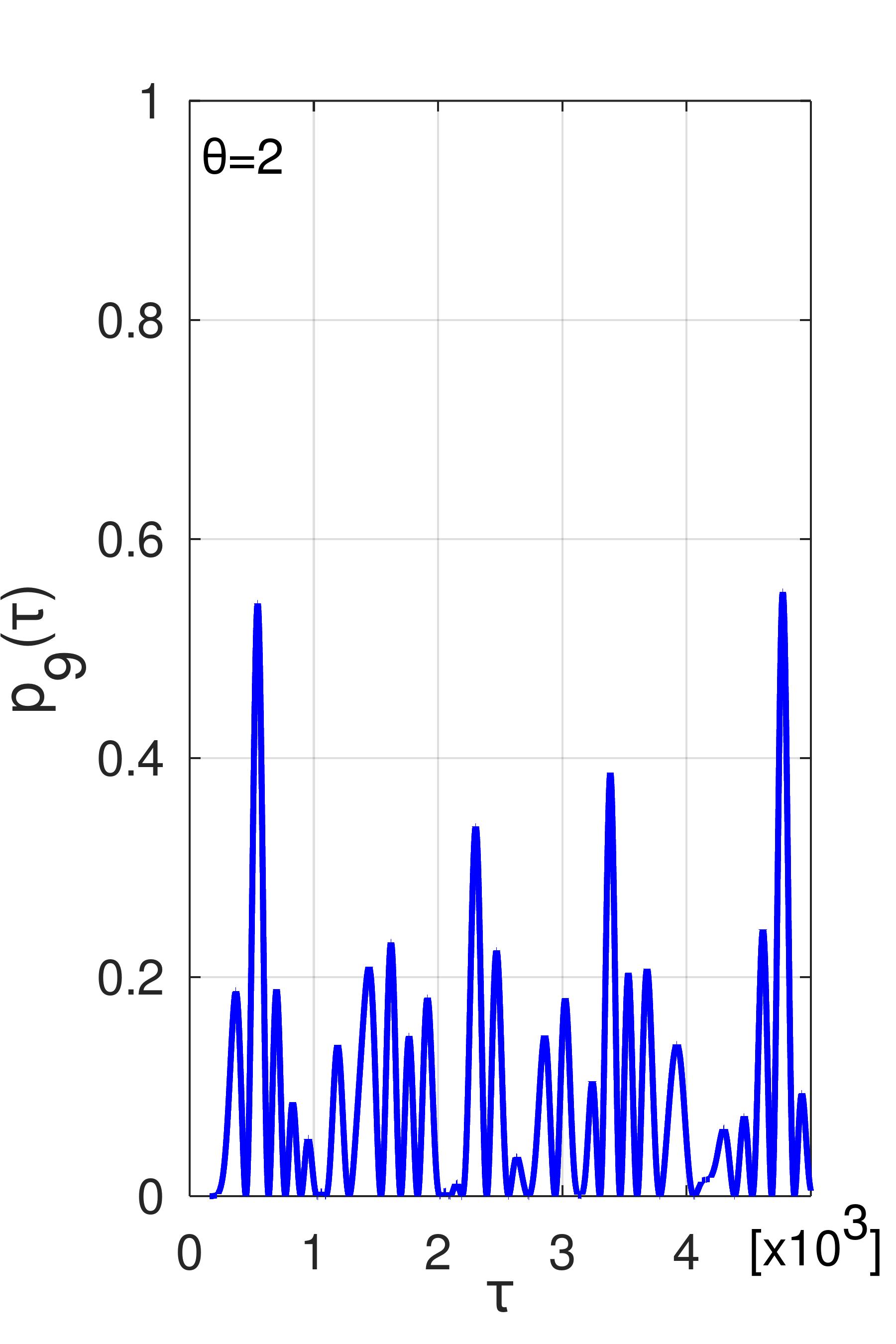}\\
			\includegraphics[width=4cm]{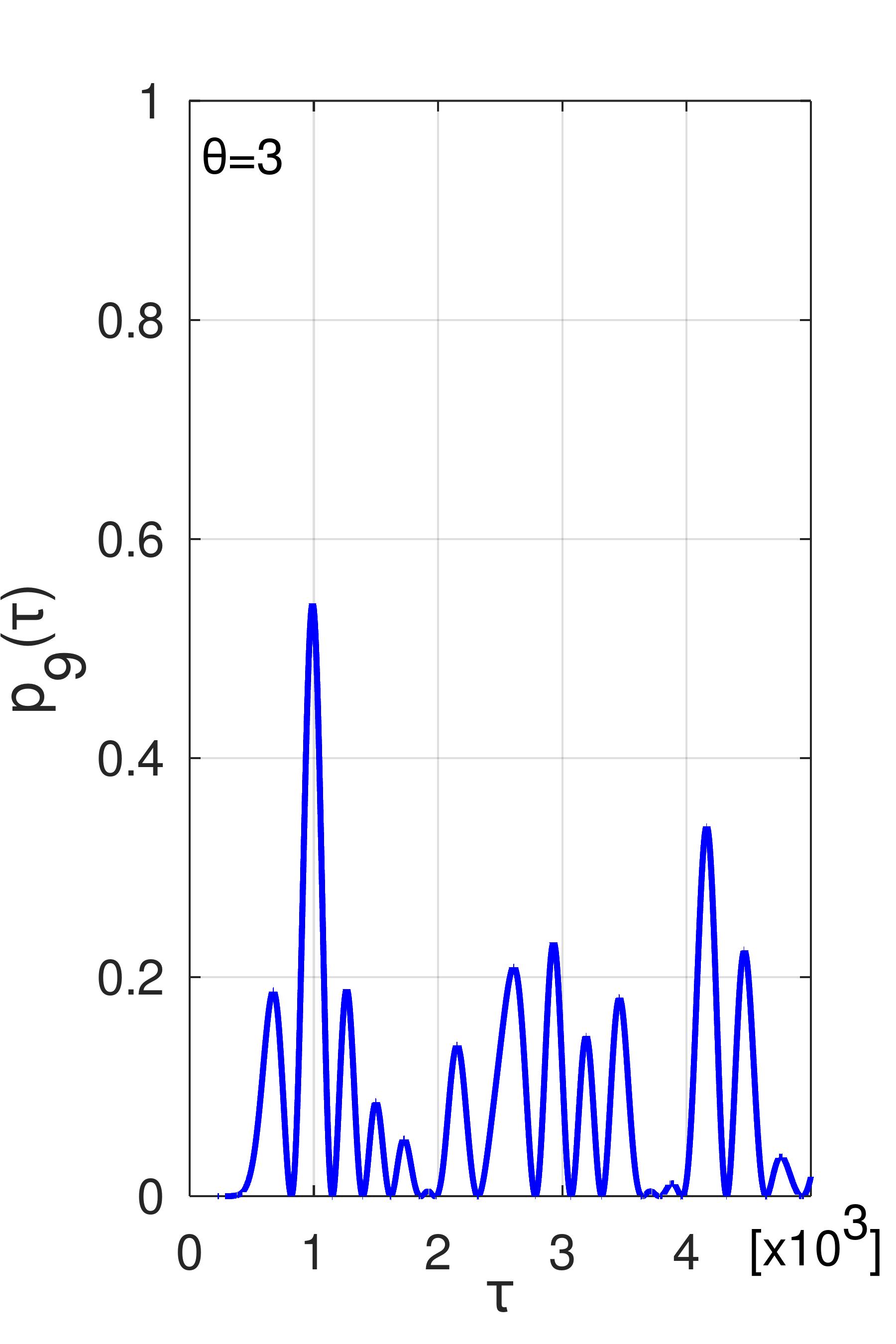}
			\includegraphics[width=4cm]{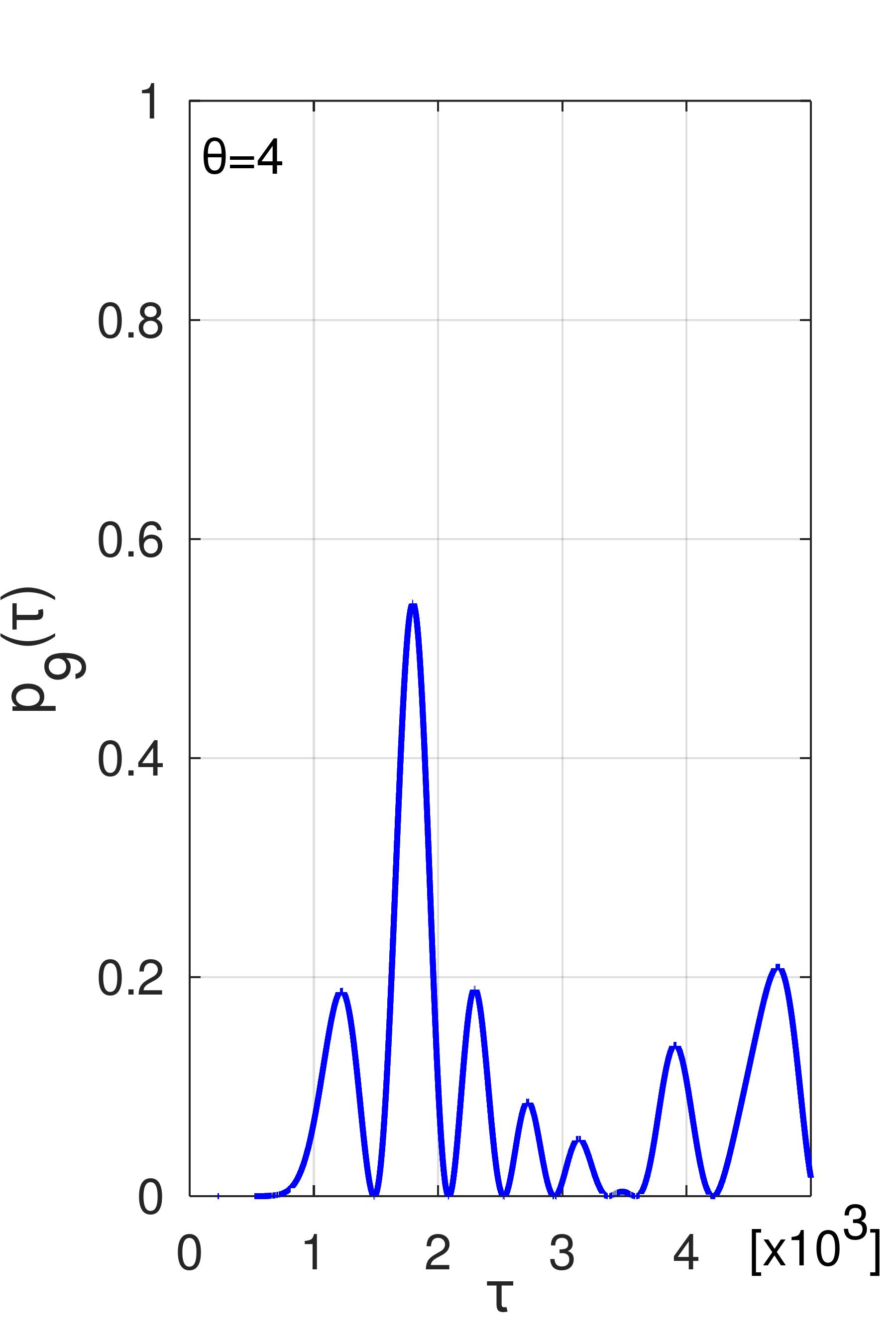} 
			\caption{Time probability distribution $p_9(\tau)=|V_9(\tau)|^2$ of excitation appearance on last ($9$th) node, for different values of system temperatures. Here, $S=0.3$ and $B=0.1$.}\label{fig03}
		\end{center}
	\end{figure}

	In Fig.\ref{fig03}, the time probability distribution $p_9(\tau)=|V_9(\tau)|^2$ for the excitation on the 9th node is shown for different system temperatures, with fixed values of \cite{AK, CevizovicPRE, Nevskaya} $\omega_0=10^{13}\;\text{s}^{-1}$. In this case, $S=0.3$, and $B=0.1$, which correspond to typical literature values for vibron excitation in $\alpha$--helix proteins.  
	
The obtained probability distributions exhibit a spectral structure composed of sequences of several peaks of comparable width grouped around one dominant maximum. The sequence closest to $\tau = 0$ corresponds to the first appearance of the excitation at the considered site after its injection into the molecular structure. The dominant peak intensity reflects the probability of finding the excitation at that site, while the recurring sequences indicate a continuous process of excitation appearance, decay, and reappearance. From Fig.\ref{fig03} it becomes evident that:\\
	
	
	
	\noindent\circnum{1} the probability of the excitation appearance at the most distant SE can be significant at all of system temperatures, meaning that the excitation has a substantial chance of appearing at a remote SE of the MC. However, increasing of $\theta$ does not affect the intensity of the probability amplitudes of the excitation appearing on remote SEs, but shifts them to higher values (slowing down the migration of the dressed quasiparticle);\\
	
	\noindent\circnum{2} the width of the probability maximum (and thus the effective "residence" time of the excitation) at the distant SE increases with the temperature of the environment. At room temperature ($\theta=4$), this residence time is on the order of $10^{-10}\;\text{s}$ or about $0.1$ ns. Such a timescale is sufficiently long to induce local modifications of the SE and, consequently, to affect the functioning of the active site to which it belongs\\. 
	
	These two findings (\circnum{1} and \circnum{2}) strongly suggest that an excitation propagating along a BmC can locally alter its properties even at sites that are far from the initial excitation point. Moreover, the excitation may remain at such remote locations long enough to influence the functionality of this chain segment in biophysical processes where it is involved. Remarkably, this effect becomes even more pronounced at temperatures characteristic of physiological conditions in living cells.\\
	
	\noindent\circnum{3} the time at which the first probability maximum appears increases with the environmental temperature, indicating that the excitation propagates more slowly as the temperature rises.\\
	
	It should be emphasized, however, that the parameter $\chi$ cannot be measured directly. For a given biomolecule and the nature of the excitation created within it, the actual value of $\chi$ depends on the theoretical model employed to evaluate this parameter from experimental data. Consequently, the true values of $\chi$ (and, therefore, of $S$) may deviate significantly from those adopted here \cite{PouthierJCP132,KalosakasPRE}.

	\subsubsection{The influence of $S$}
	
\begin{figure}[h]
		\begin{center}
			\includegraphics[width=4cm]{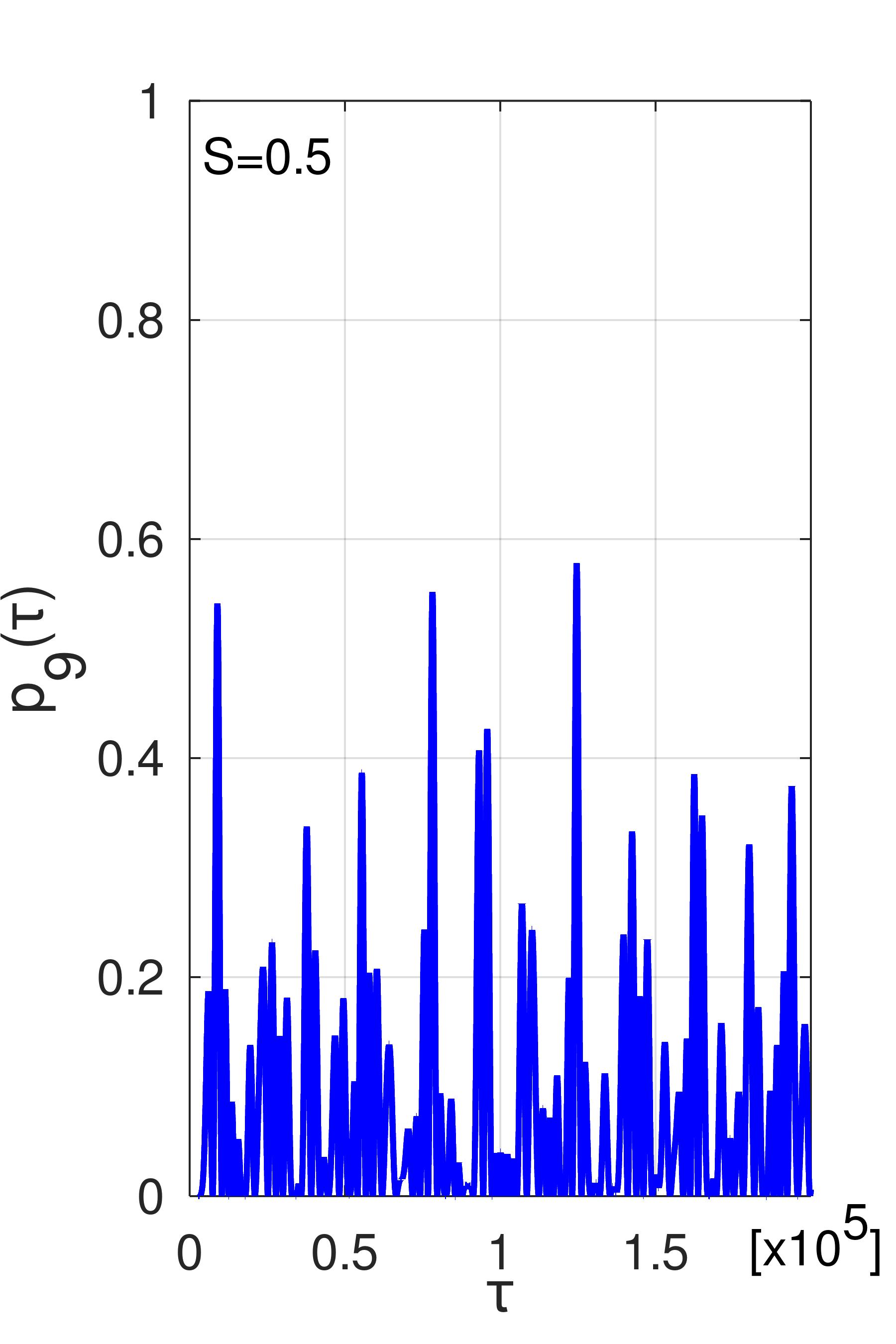}
			\includegraphics[width=4cm]{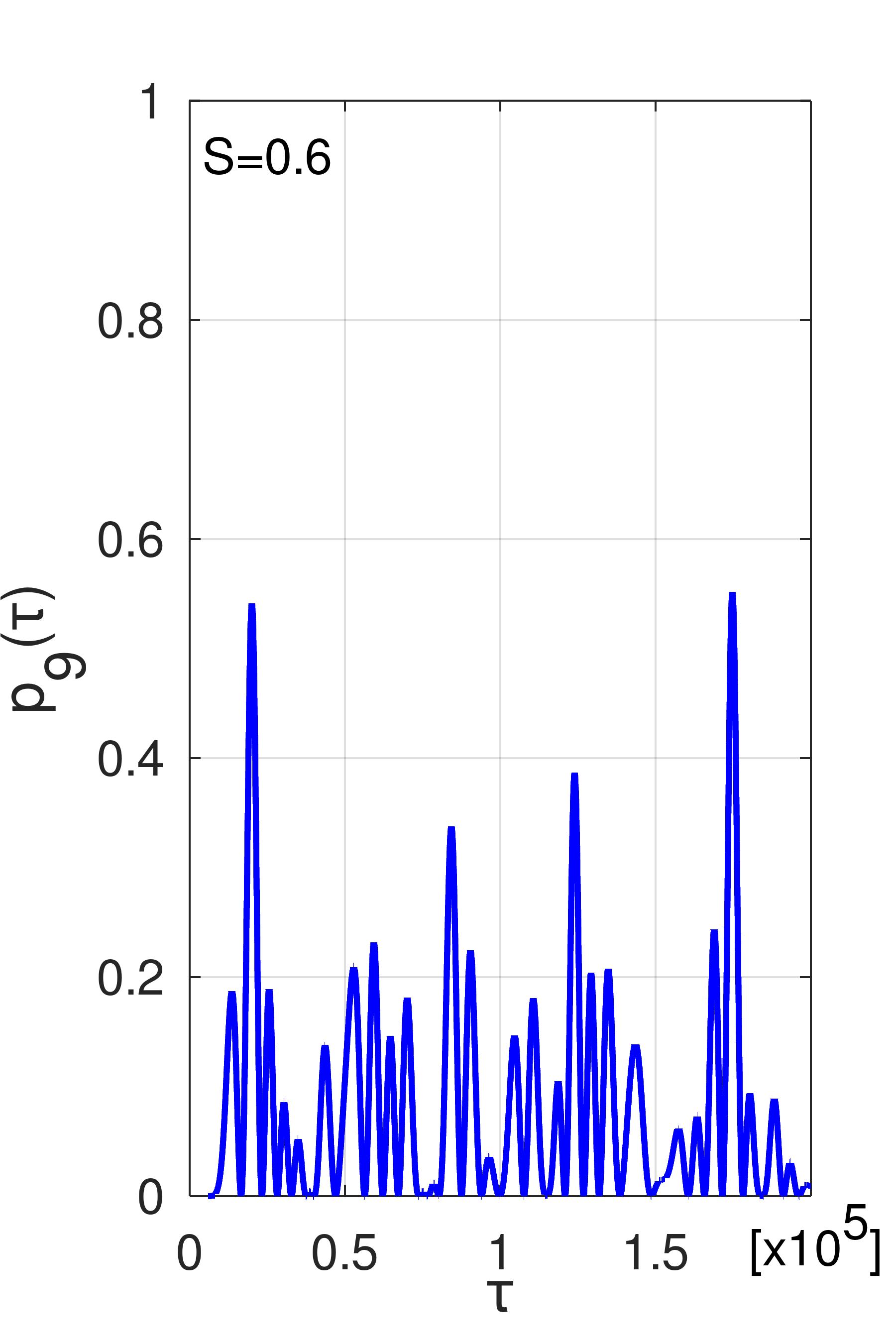}\\
			\includegraphics[width=4cm]{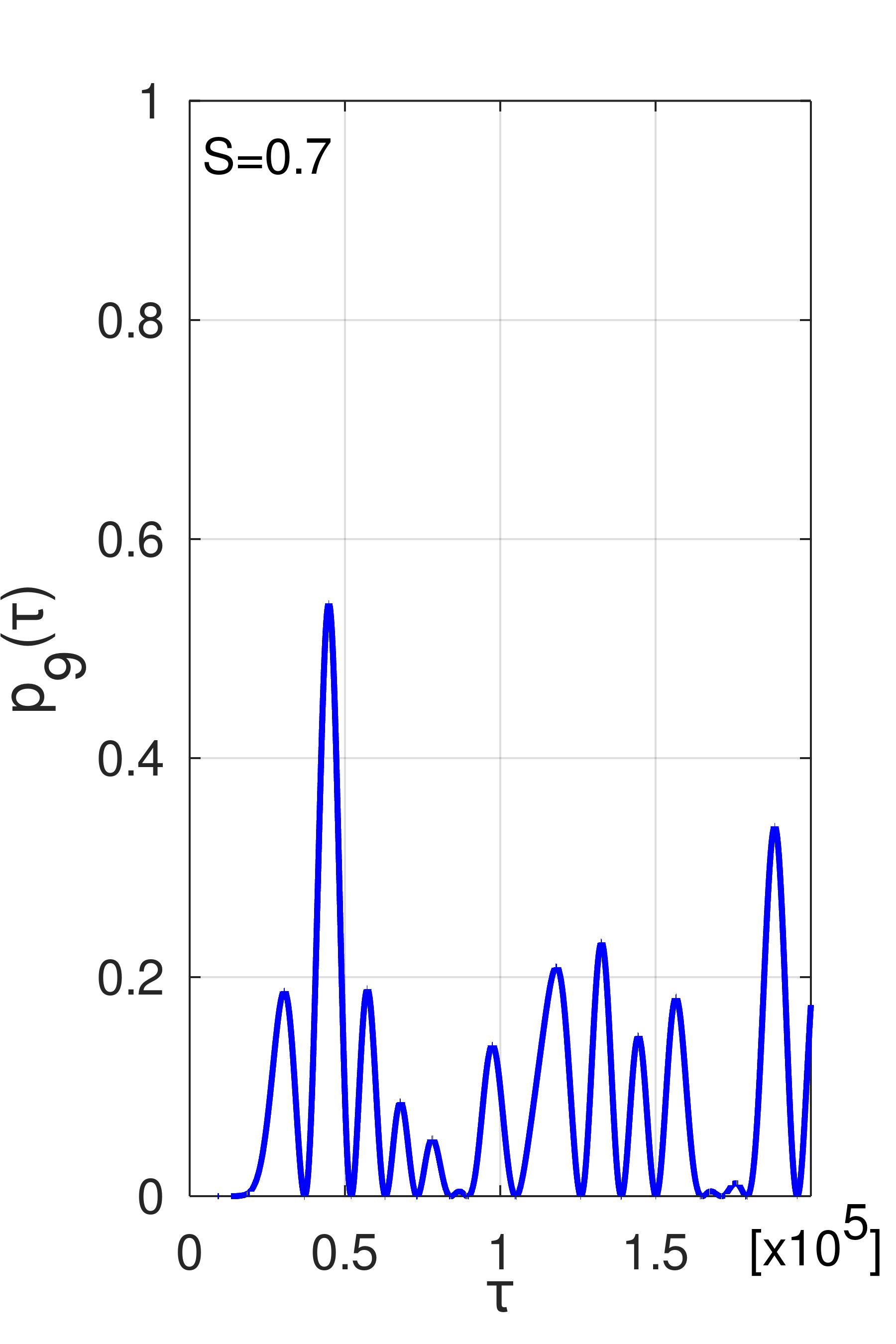}
			\includegraphics[width=4cm]{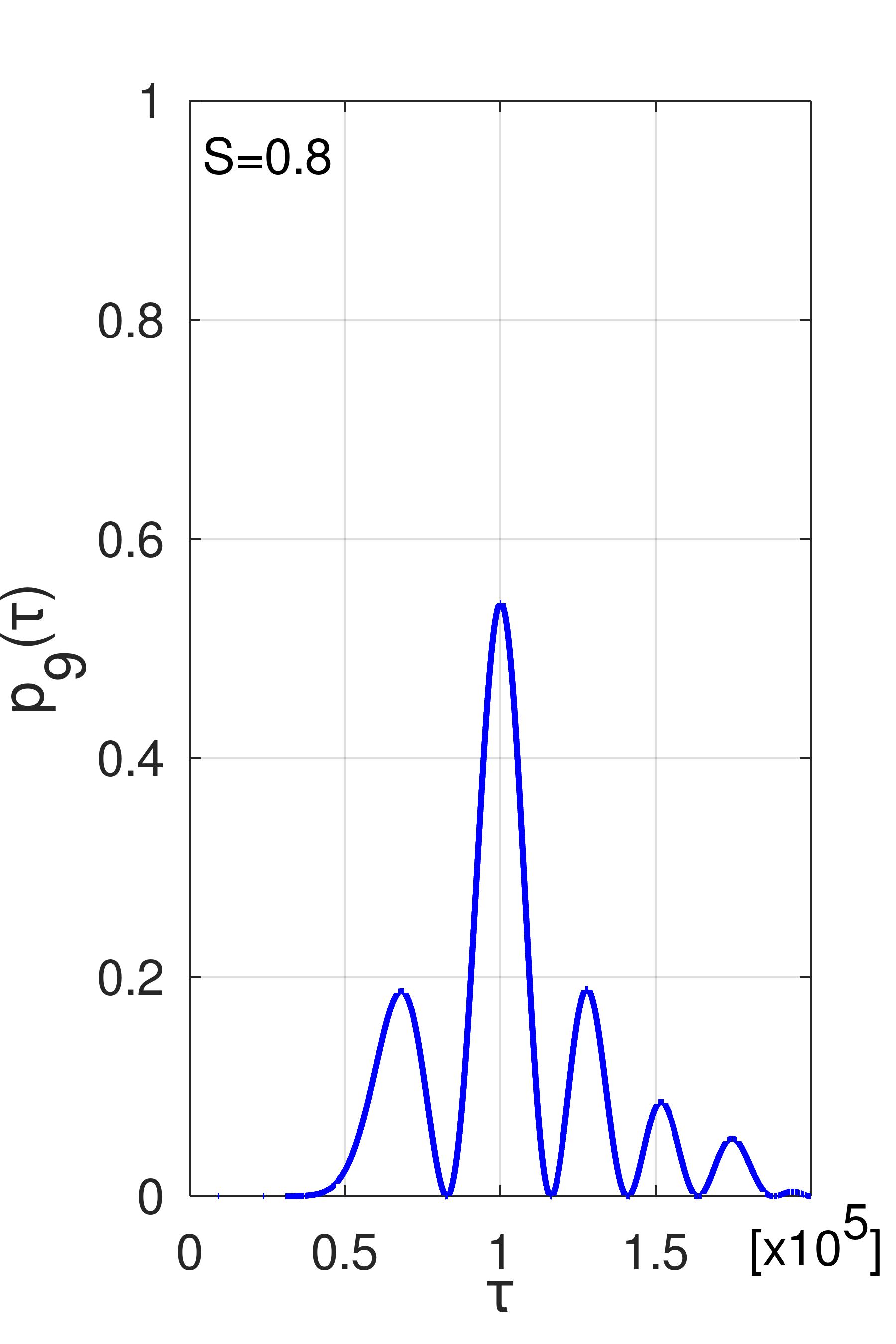} 
			\caption{Time probability distribution $p_9(\tau)=|V_9(\tau)|^2$ of excitation appearance on last ($9$th) node, for different values $S$. Here, $B=0.1$ and $\theta=4$.}\label{fig04}
		\end{center}
	\end{figure}

	As already mentioned, the value of the parameter $\chi$, which determines the strength of the interaction between the excitation and the thermal oscillations of the molecule, is not precisely known. In a preliminary analysis of Eqs.(\ref{V9tSB2}) and (\ref{Omega}), we concluded that an increase in the parameter $S$ leads to an increase in the residence time, while the amplitudes of the probability remain essentially unaffected. We now turn to a more detailed investigation of this effect. Specifically, we analyze how small variations in $S$ influence both the probability of finding the excitation at a given structural element of the molecular chain and the corresponding residence time at that SE. In Fig.\ref{fig04}, we present the time distribution of the probability $p_9(\tau)=|V_9(\tau)|^2$ of excitation at the 9th SE, for fixed values of $B$ and temperature $\tau$, and for different values of $S$. The excitation is initially induced at the second SE, located on the left side of the MC.

	From Fig.\ref{fig04} we see that:\\
	
	\noindent\circnum{1} Similar to the influence of the environmental temperature, increasing $S$ does not affect the intensity of the probability maxima, but shifts them to larger times, i.e., it slows down the migration of the dressed quasiparticle;\\
	
	\noindent\circnum{2} with increasing values of the parameter $S$, the residence time of the excitation at the distant SE becomes significantly longer. For instance, at $S=0.3$ and $\theta=4$ (Fig.\ref{fig03}) the width of the first maximum is on the order of $0.05$ ns, while at $S=0.8$ and $\theta=4$ (Fig.\ref{fig04}) it reaches $\sim 4$ ns, implying that the residence time of the excitation at the last node is approximately 80 times longer;\\
	 
	\noindent\circnum{3} we further observe that for $S=0.3$ and $\theta=4$ (Fig.\ref{fig03}), the first maximum at the last SE of the MC appears at $\sim 0.05$ ns, whereas for $S=0.8$ and $\theta=4$ (Fig.\ref{fig04}), the excitation requires $\sim 10$ ns to reach the last SE.\\
	
	The presented results confirm and further refine the conclusions drawn from the analysis of Eqs. (\ref{V9tSB2}) and (\ref{Omega}). In particular, within the framework of the proposed model, temperature and coupling strength have no effect on the maxima of the probabilities. However, they significantly influence both the width and the position of these maxima, which is essential for further investigations of physiological and biological systems under realistic conditions.

	\subsection{How does the length of MC affect the $p_n({\tau})$?}
	
\begin{figure}[h]
			\begin{center}
				\includegraphics[width=4cm]{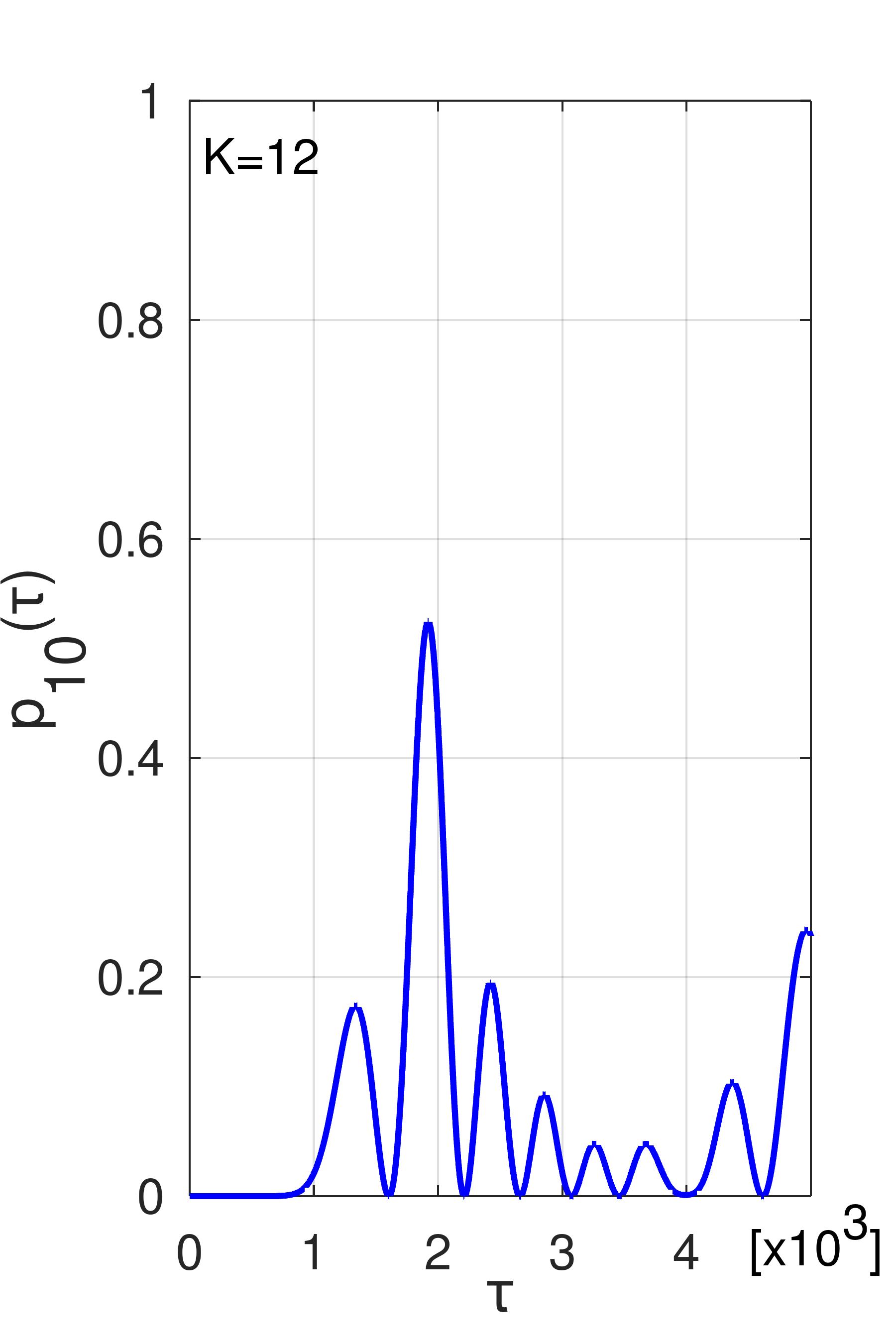}
				\includegraphics[width=4cm]{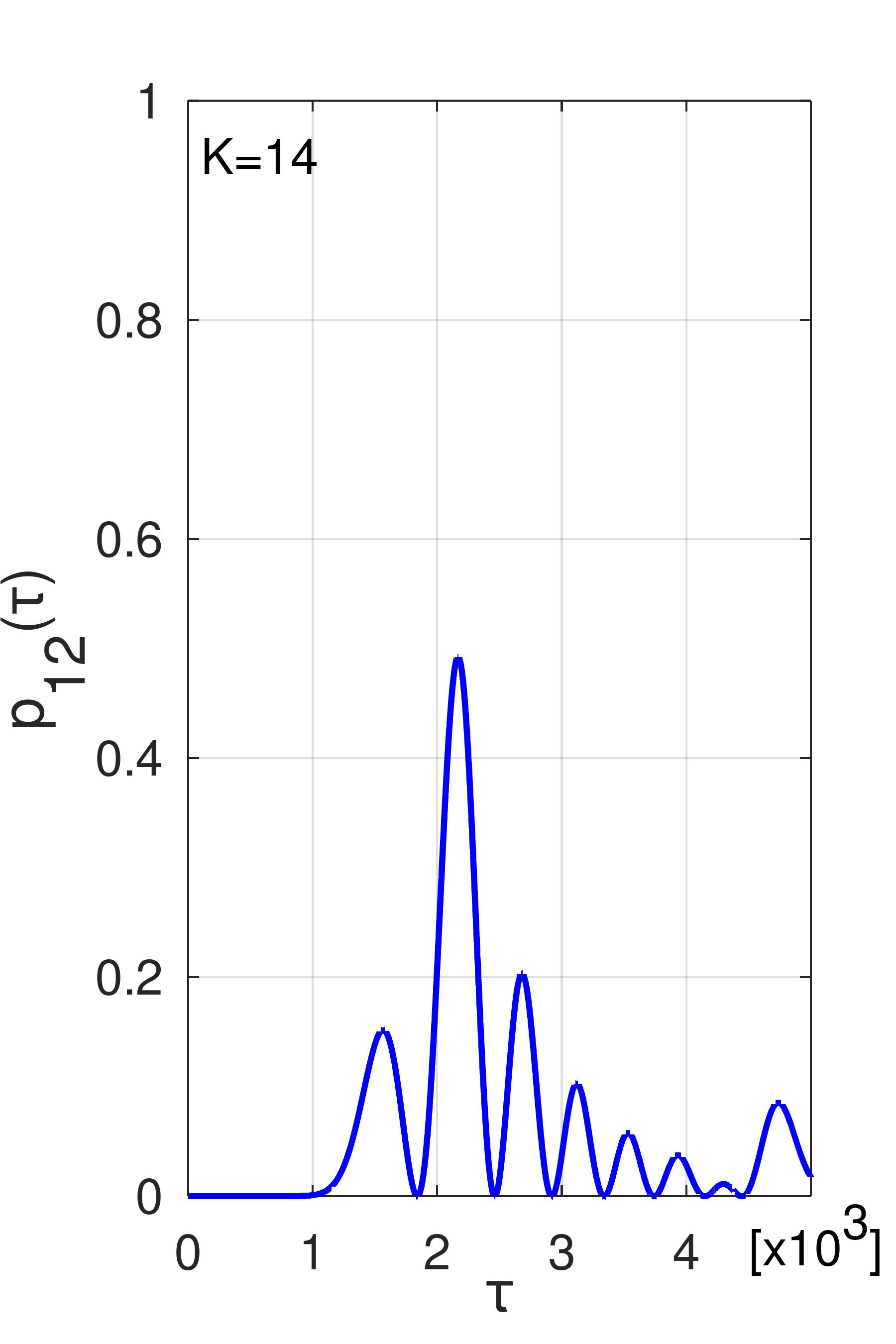}\\
				\includegraphics[width=4cm]{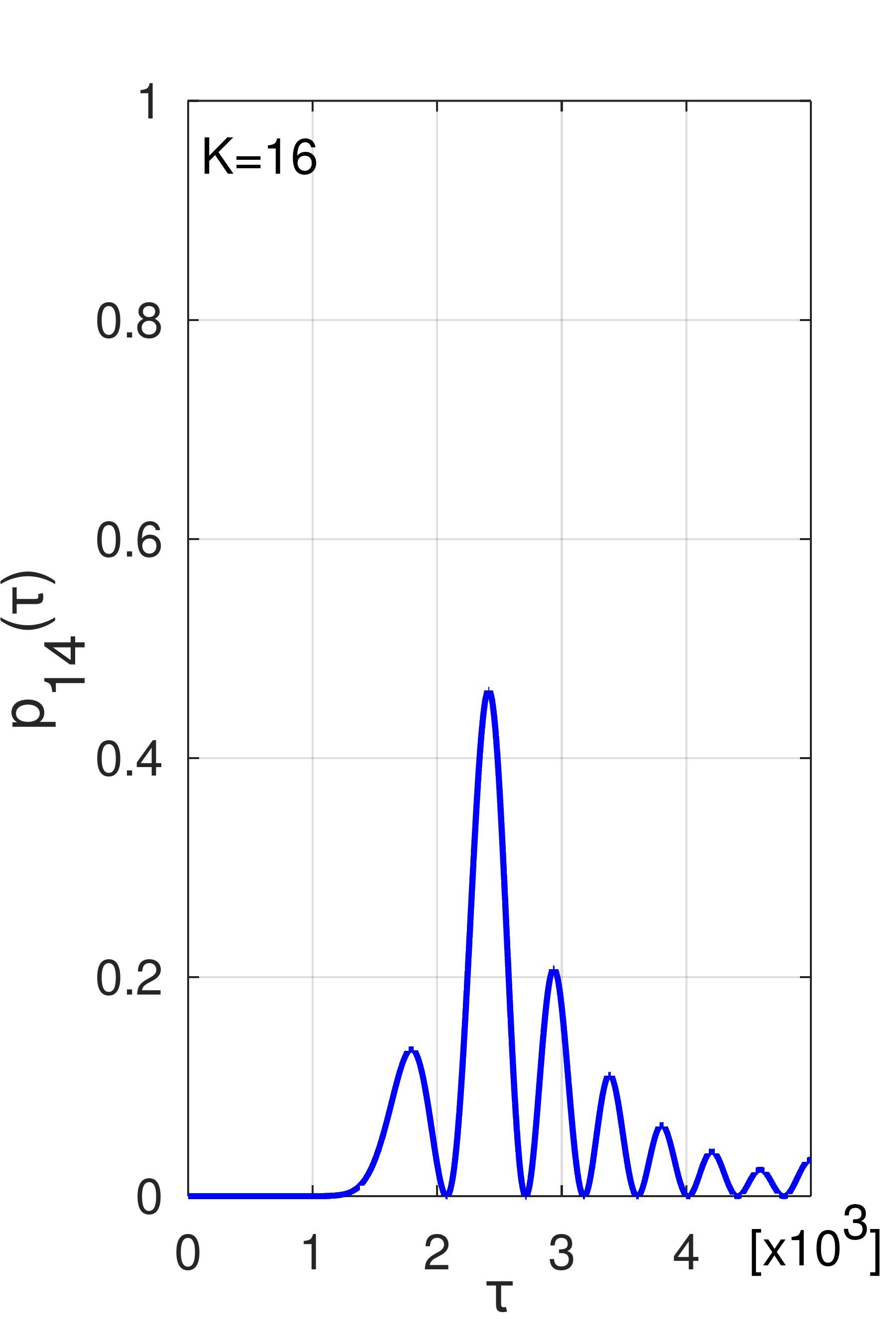}
				\includegraphics[width=4cm]{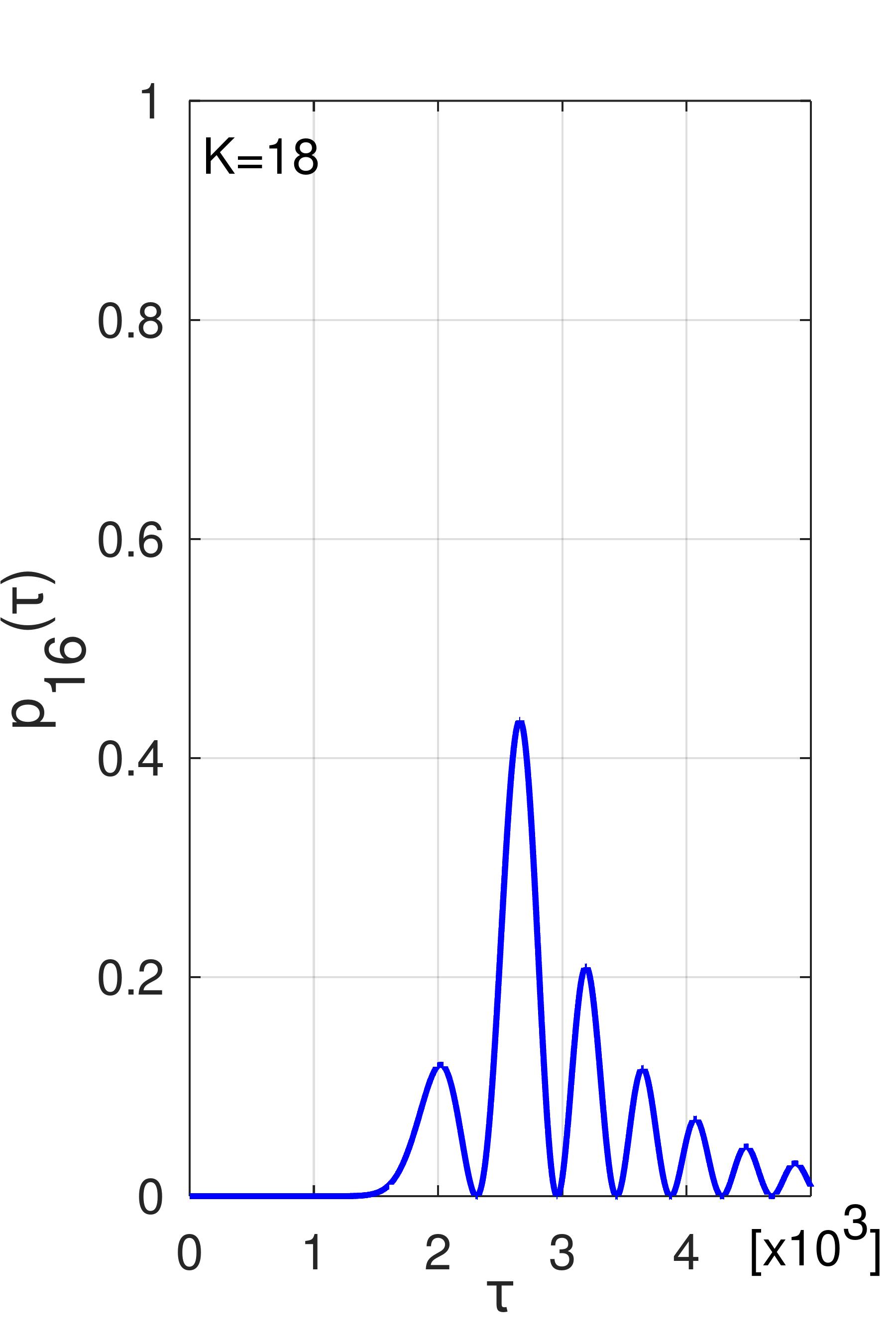}
				\caption{Time distribution of the excitation probability on the last node of the MC. Parameters: $S=0.3$, $B=0.1$, $\theta=4$. Excitation was induced at the second SE from the left.}\label{fig05}
			\end{center}
		\end{figure}

	Let us now consider how the length of the MC affects the probability distribution of finding the excitation at the distant nodes of the structure. In Eqs.(\ref{V9tSB2}) and (\ref{Omega}), this effect is reflected in the order of the polynomials $D_n(x)$ and their roots, which determine the intensity and width of the probability maxima. However, it is difficult to draw concrete conclusions from these equations alone. Therefore, we consider several configurations of MCs of different lengths, consisting of $K=N+M+1 = \{12,14,16,18\}$ SEs, with the excitation initially induced at the second element on the left side of the chain in all cases. In Fig.\ref{fig05}, we present the time distribution of the probability $p_n(\tau)=|V_n(\tau)|^2$ of finding the excitation at the most distant node for chains of different lengths at room temperature ($\theta=4$).
	
It can be observed that, with increasing molecular chain length, the probability of finding the excitation at the last node slightly decreases, while its residence time remains unchanged. The time required for the excitation to reach it increases with the chain length.

	\subsection{How does the position of the initially excited node affect $p_n(\tau)$?}

\begin{figure}[H]
		\begin{center}
			\includegraphics[width=8cm]{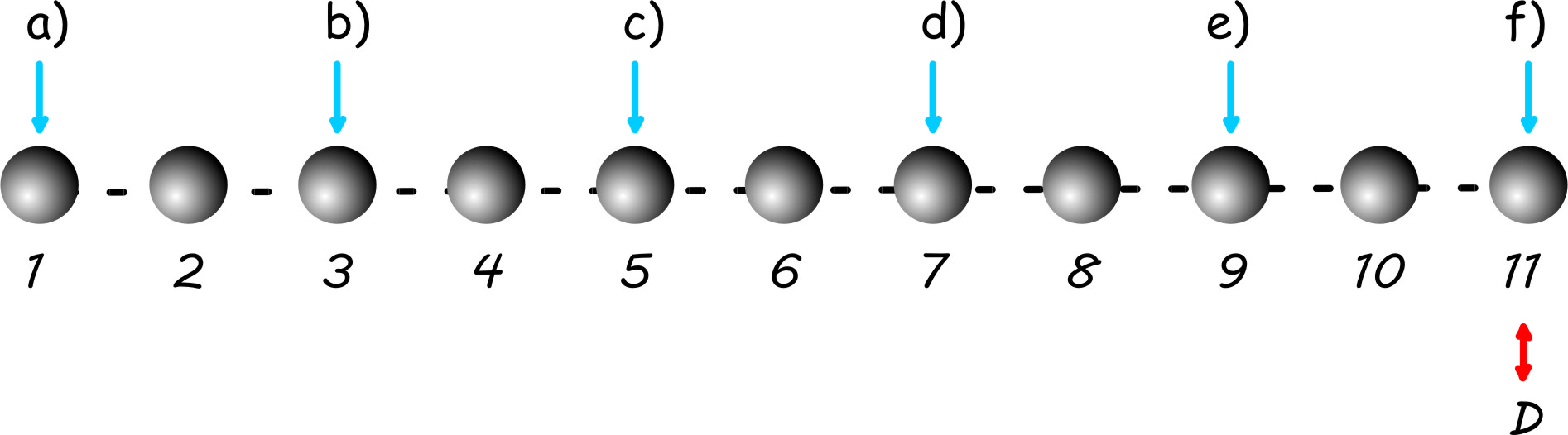}
			\caption{Schematic of the MC showing the initially excited node (arrows above nodes) and the node “D” where the probability of the excitation appearance is calculated.}\label{fig07}
		\end{center}
	\end{figure}

	Finally, we examine how the probability of finding the excitation at a particular node of a finite MC depends on the position where the excitation is initially induced. For that purpose, we considered a MC consisting of 11 SEs and analyzed the probability of finding the excitation at the far--right SE, denoted schematically as “D” (detection) in Fig.\ref{fig07}. The initially excited node is indicated by an arrow above the chain, which also points to the corresponding panel in Fig.\ref{fig08}, showing the time evolution of the probability at node “D”.
	
	\noindent Obtained results are presented on Fig.(\ref{fig08}):
	
	\begin{widetext}
		\begin{figure*}[htb]
			\begin{center}
				\includegraphics[width=4cm]{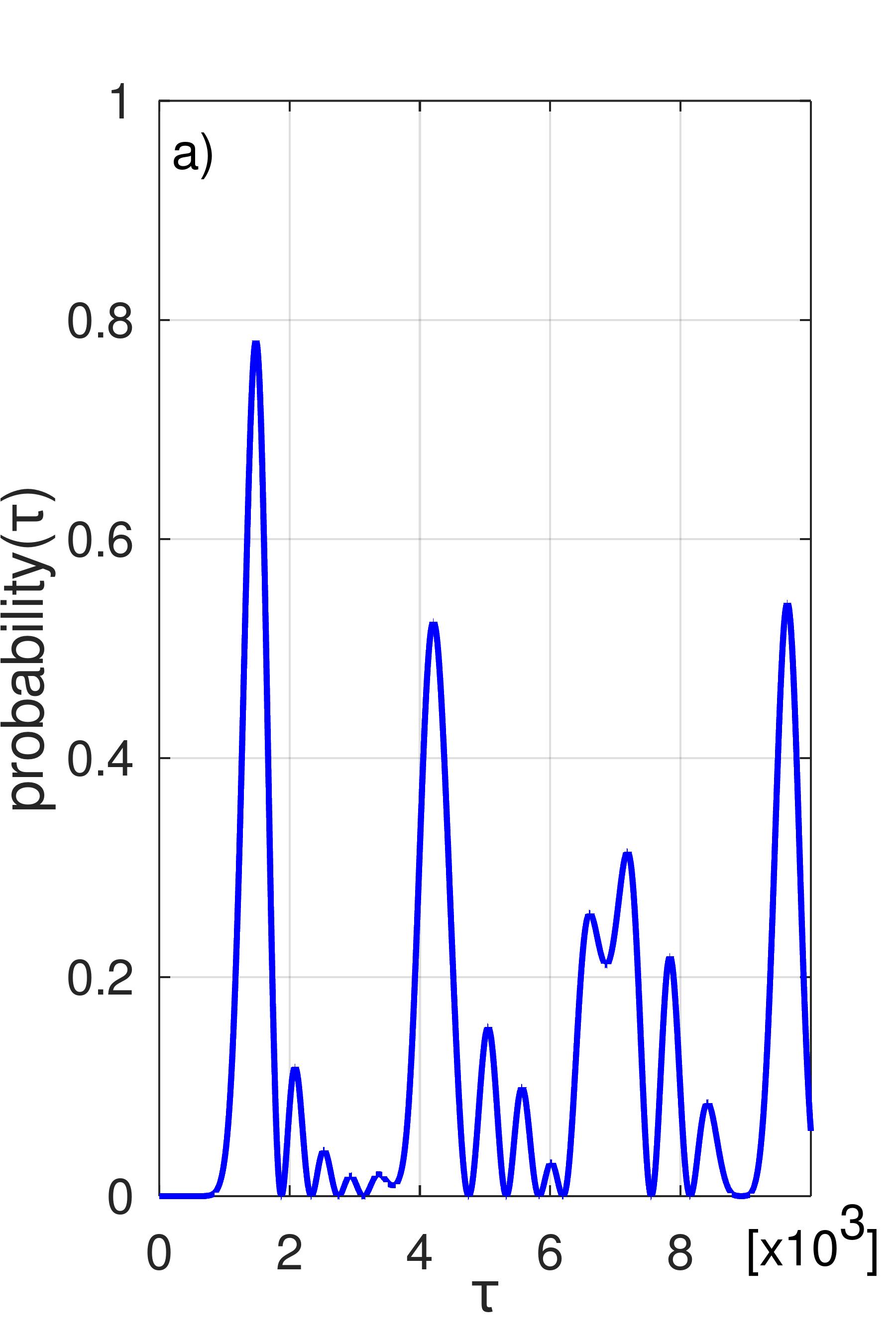}
				\includegraphics[width=4cm]{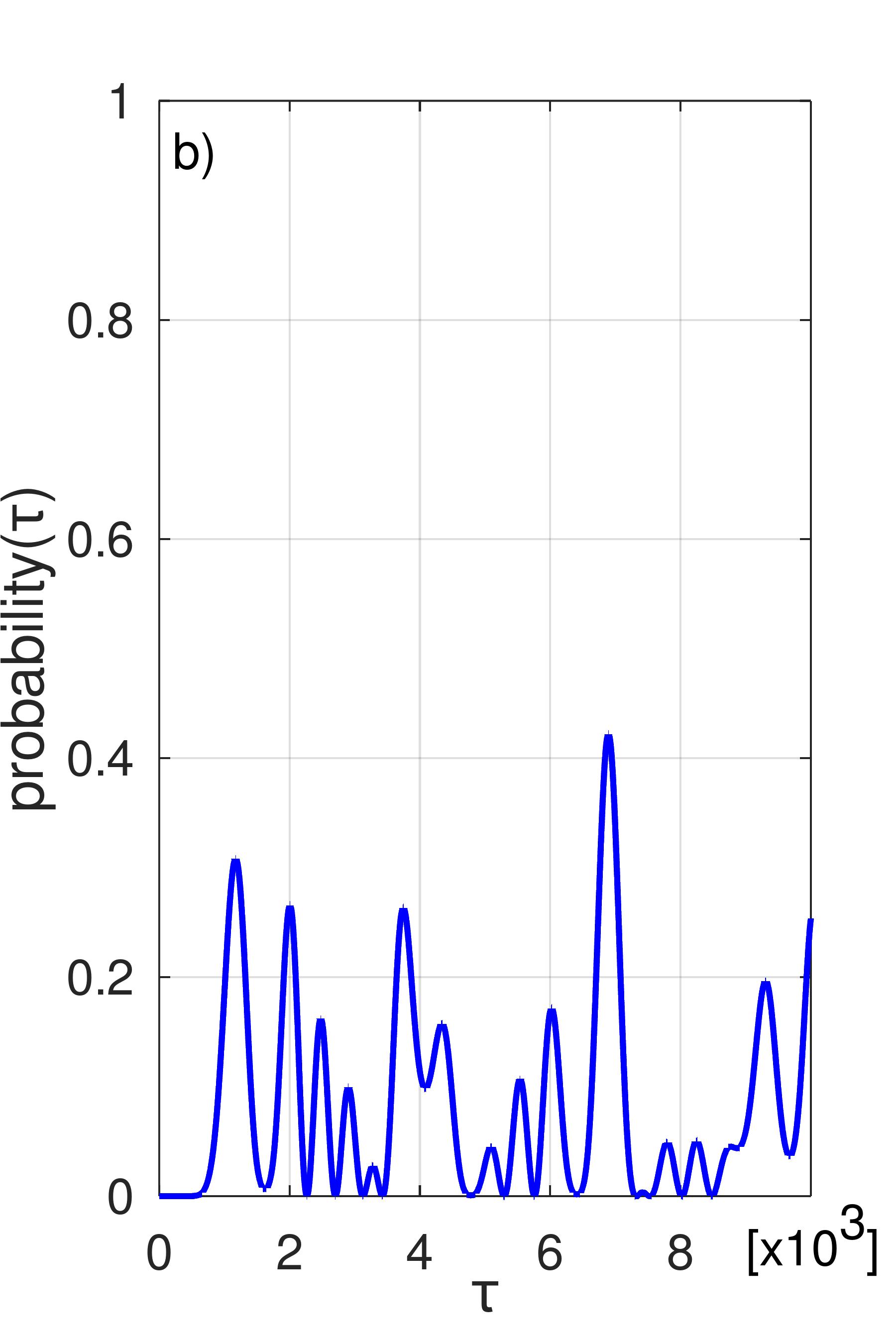}
				\includegraphics[width=4cm]{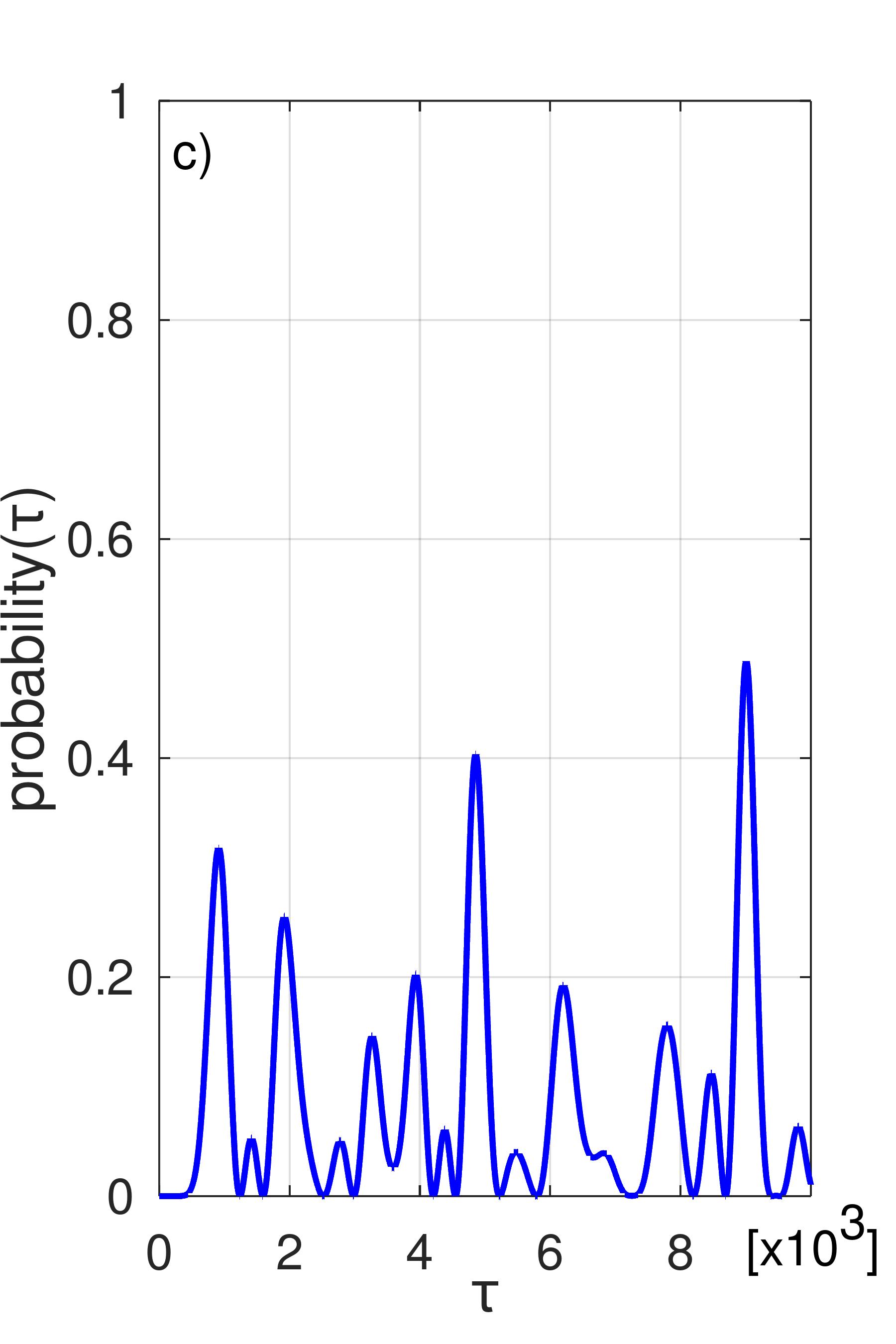}\\
				\includegraphics[width=4cm]{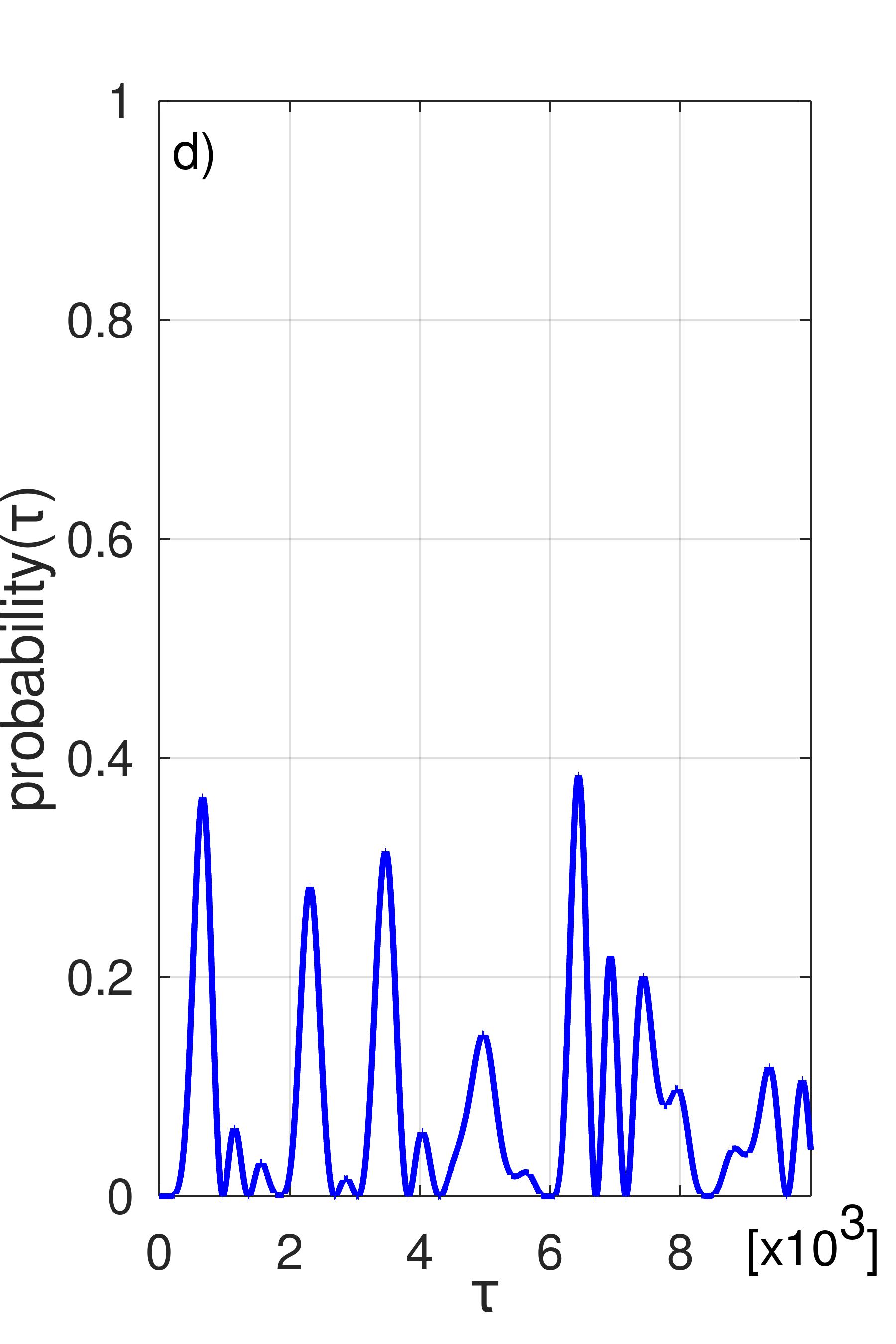}
				\includegraphics[width=4cm]{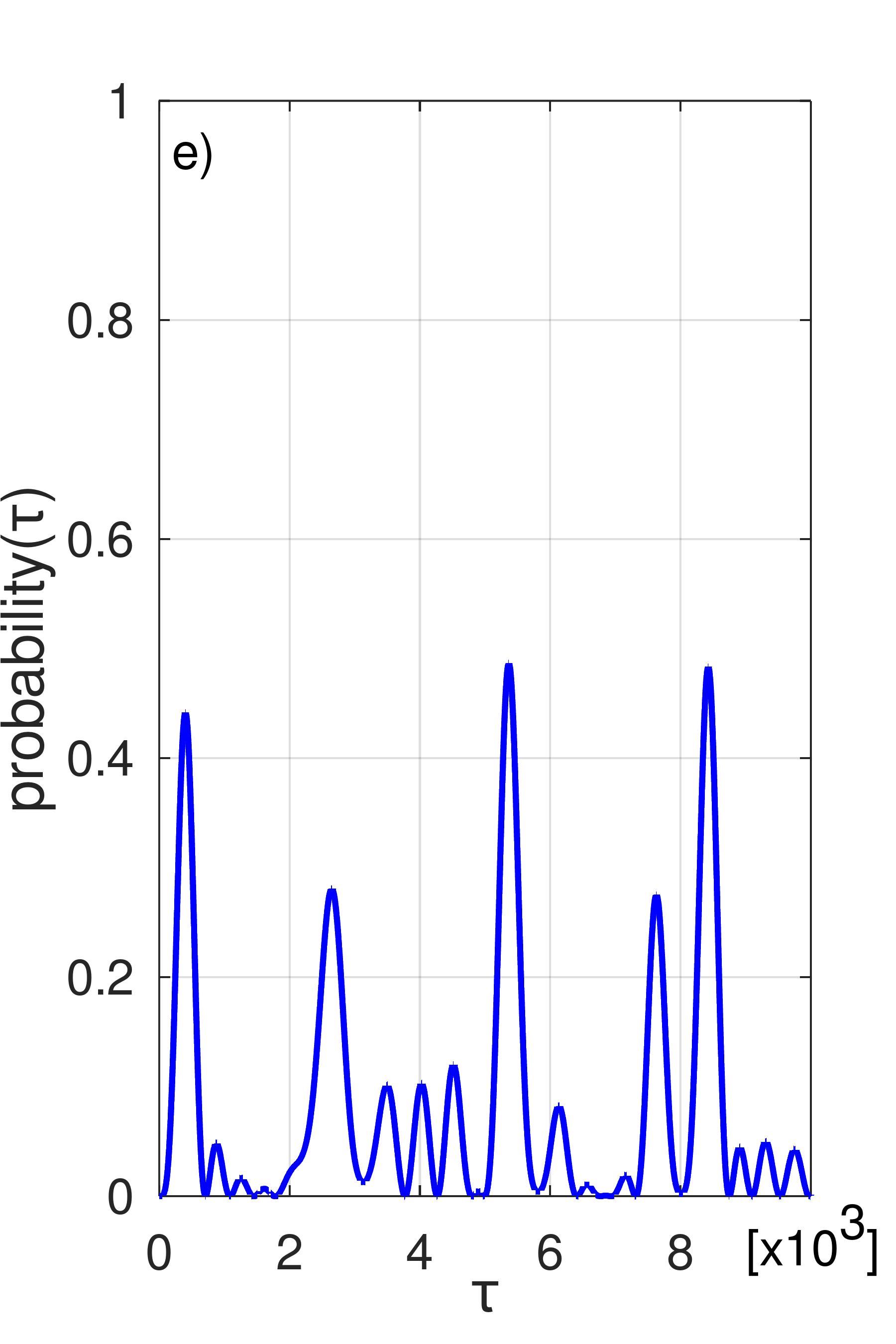}
				\includegraphics[width=4cm]{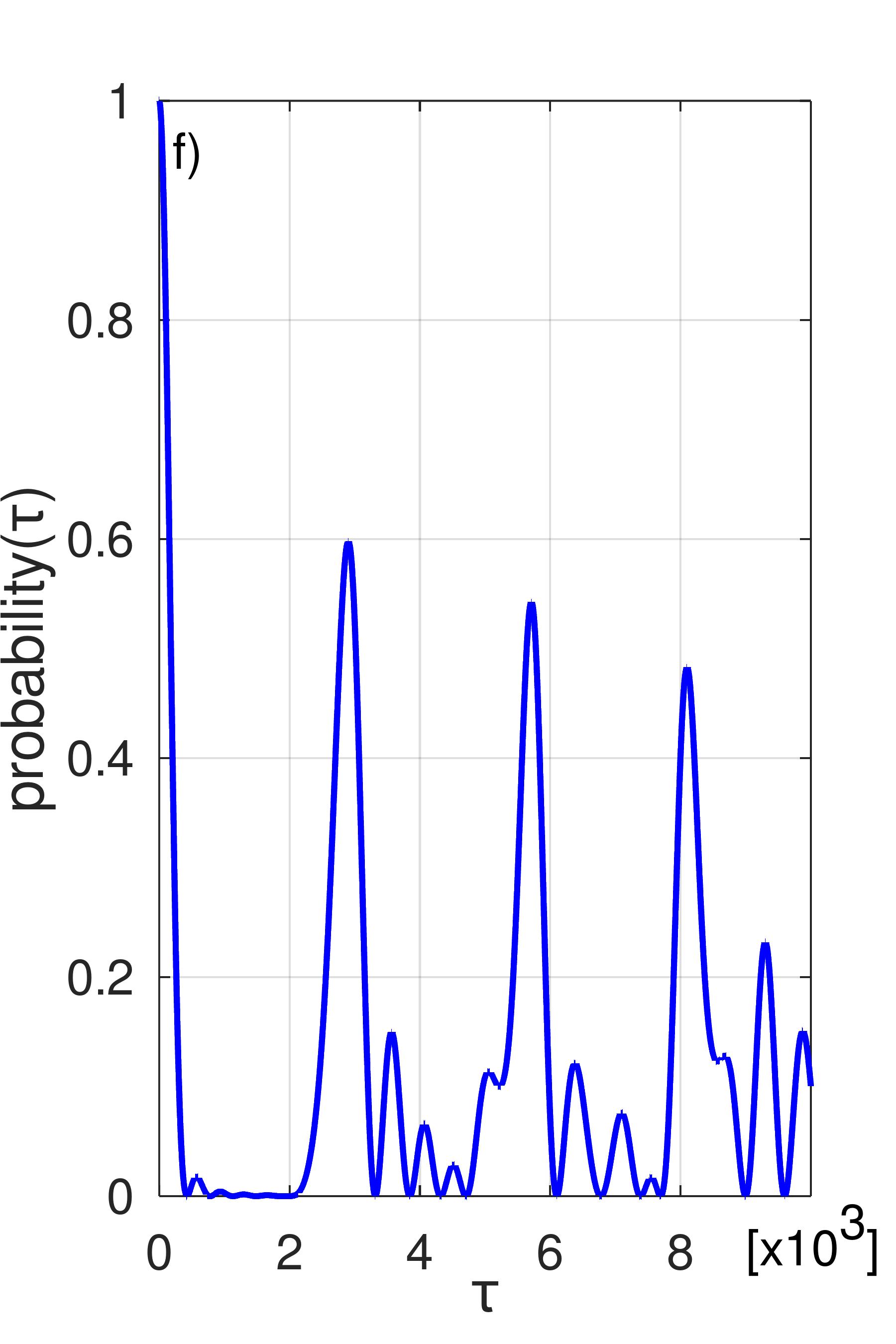}
				\caption{Time distribution of the probability of excitation on the last SE on the right side. Here, $S=0.3$, $B=0.1$ and $\theta=4$.}\label{fig08}
			\end{center}
		\end{figure*}
	\end{widetext}
	
By varying the initially excited node, the temporal probability distributions at the last node change significantly. Considering the first significant peak in these distributions, the highest probability of finding the excitation at the last node occurs when the excitation is initialy induced at the first node on the left. Small changes in the position of the initial excitation site strongly reduce the peak intensity at the last node, while the peak width, reflecting the residence time, remains unchanged. These observations suggest the possibility of controlling excitation migration along the molecular chain by selecting the initial excitation site (targeting).

	\section{Conclusion}
	
		The results highlight excitation migration as a potential cause of disruption of BmC functions during physiological processes within a living cell. An excitation induced in a BmC lacking sufficient energy to destroy its structure may cause local changes in its physical properties, such as charge density and electric dipole distribution, thereby affecting its physiological functions. This effect depends on both the probability of excitation occurring at a given node and the residence time of the excitation at that node. Moreover, changes in the physical state of a local segment of the BmC are not necessarily confined to the site of the initial excitation: due to resonant interactions between neighboring SEs, the excitation may appear at practically any node of the BmC, including distant ones. The main findings of the presented model can be summarized as follows:
	
	\begin{enumerate}
		\item The probability of excitation occurrence can be significant at nodes distant from the initially excited node. In addition, the residence time of the excitation on SEs of the BmC can be sufficiently long for the affected chain segment to exhibit altered functionality in physiological processes.
		
		\item The temperature of the environment does not influence the magnitude of the probability, but it affects both the propagation time of the excitation through the BmC (slowing down as the temperature increases) and the residence time of the excitation at a given node of the BmC.
		
		\item The residence time of an excitation at a given node increases with both the environmental temperature and its coupling to the thermal oscillations of the BmC. At temperatures relevant for physiological processes in living cells, the induced excitation can persist long enough to alter the functionality of a local segment of the BmC.
		
		\item The position of the initially excited node in a finite--length BmC affects the probability distribution of the excitation across the chain. As this node shifts toward the center of the molecular chain, the probability of finding the excitation at the terminal node slightly decreases, whereas its residence time remains unchanged.
	\end{enumerate}

It should be noted that the appearance of the excitation at nodes distant from the initially excited one is a purely quantum effect, resulting from the manifestation of quantum coherence within a molecular structure of finite length. In this regard, it is interesting to examine the probability of finding the excitation on both sides of the initially excited node and to compare the corresponding probability distributions. Specifically, any asymmetry between these distributions would indicate that the position of the node where the excitation is induced, together with the preservation of the excitation's quantum nature, can lead to a preference for one of the two possible directions of excitation migration along a structure composed of identical structural elements. This effect is similar to quantum ratcheting, except that here it does not arise from an asymmetry in the physical properties of the system (such as an asymmetric potential, as in the “standard” quantum ratcheting case). It provides a framework for understanding directed motion of quantum excitation even in completely homogeneous and spatially confined structures. The consideration of this effect represents the next stage of our investigation. Furthermore, the derived model can be used to estimate poorly known physical parameters of BmCs. For instance, by measuring the probability of vibron or electronic excitations at different SEs, the model allows estimation of vibron--phonon and electron--phonon interaction constants, which remain largely unknown for various polypeptide chains. In addition, it can provide a theoretical foundation for developing models and technologies aimed at correcting errors in damaged biomolecules and restoring their altered functions.
	

	\begin{acknowledgments}
		This work was supported by the Ministry of Science, Technological Development, and Innovation of the Republic of Serbia through the Project contract No 451-03-136/2025-03/200017.\\
	\end{acknowledgments}

	\appendix
	\section{Solving the system of equations for Laplace transformations of the correlation functions $\tilde{V}_n(x)$}
	
	By introducing the auxiliary variable $x$ through Eq.(\ref{x}) where $x\in \mathbb{C}$, the system of equations given in Eq.(\ref{dVlindesno}) can be rewritten as:
	
	\begin{align}\label{dVlin3bb}
	x\tilde{V}_1&=\tilde{V}_0+\tilde{V}_2\nonumber\\
	......&...................\nonumber\\
	x\tilde{V}_{N-2}&=\tilde{V}_{N-3}+\tilde{V}_{N-1}\\
	x\tilde{V}_{N-1}&=\tilde{V}_{N-2}+\tilde{V}_N\nonumber\\
	x\tilde{V}_N&=\tilde{V}_{N-1}\nonumber
	\end{align}
	
	\noindent The resulting system of coupled algebraic equations can be solved successively, starting from the last equation. From the final equation, we express $\bar{V}_N$ in terms of $\bar{V}_{N-1}$ $$\tilde{V}_N=\frac{1}{x}\tilde{V}_{N-1}=b_1\tilde{V}_{N-1}$$ where $ b_1=\frac{1}{x}$. We then substitute this expression for $\tilde{V}_N$ into the previous equation from Eq.(\ref{dVlin3bb}), establishing a relation between $\bar{V}_{N-1}$ and $\bar{V}_{N-2}$:  $$\tilde{V}_{N-1}=\frac{1}{x-\frac{1}{x}}\tilde{V}_{N-2}=\frac{1}{x-b_1}\tilde{V}_{N-2}=b_2\tilde{V}_{N-2}$$ where $b_2=\frac{1}{x-b_1}$, then $$\tilde{V}_{N-2}=\frac{1}{x-\frac{1}{x-\frac{1}{x}}}\tilde{V}_{N-3}=\frac{1}{x-b_2}\tilde{V}_{N-3}=b_3\tilde{V}_{N-3}$$ where $b_3=\frac{1}{x-b_2}$. By repeating the procedure successively, we relate $\bar{V}_{N-\kappa}$ to $\bar{V}_{N-\kappa-1}$: $$\tilde{V}_{N-\kappa}=b_{\kappa+1}\tilde{V}_{N-\kappa-1}$$ where $b_{\kappa+1}=\frac{1}{x-b_{\kappa}}$. This continues up to $\kappa=N-2$ $$\tilde{V}_2=b_{N-1}\tilde{V}_1$$ and finally for $\kappa=N-1$  $$\tilde{V}_1=b_N\tilde{V}_0$$ In this way, the LT of the CF at any SE of the MC can be expressed in terms of the LT of the CF at the preceding node.

	
	In the next step, by successively substituting $\tilde{V}_1$ into $\tilde{V}_2$, we express $\tilde{V}_2$ in terms of $\tilde{V}_0$. Repeating this procedure, we obtain the LT of the CF at any SE of the MC as a function of the LT of the CF at the zeroth node, i.e., $\tilde{V}_0$. Finally, we obtain:
	
	\begin{align}
	\tilde{V}_1&=b_N\tilde{V}_0\nonumber\\
	\tilde{V}_2&=b_{N-1}b_N\tilde{V}_0\nonumber\\
	...&.............................\nonumber \\
	\tilde{V}_k&=b_{N-k+1}b_{N-k+2}\cdot ...\cdot b_N\tilde{V}_0;\quad k\in\left\{2,3,...,N-1\right\}\nonumber\\
	...&.............................\nonumber\\
	\tilde{V}_N&=b_1b_2\cdot ...\cdot b_N\tilde{V}_0\nonumber
	\end{align}
	
	\noindent The scheme for solving the system of equations for the Laplace transforms of the correlation functions is graphically illustrated in Fig.\ref{fig09}:
	
	\begin{figure}[H] 
		\begin{center}
			\includegraphics[width=80mm]{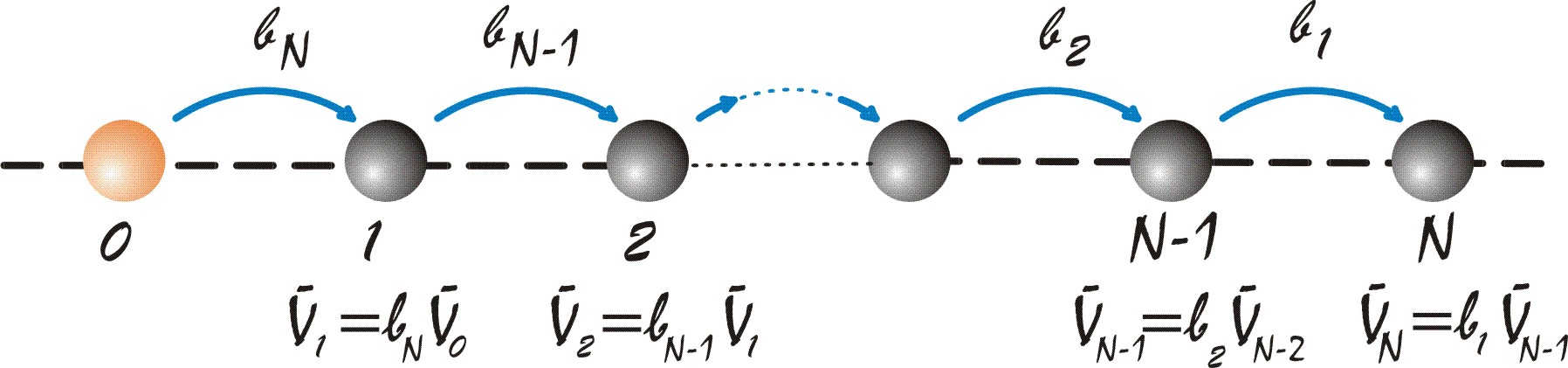}
			\caption{Scheme for solving the system of equations for the Laplace transforms of the correlation functions.}\label{fig09}
		\end{center}
	\end{figure}

	\noindent The coefficients $b_1,b_2,\ldots,b_N$ are generated according to the following rule: 
	
	\begin{equation}\label{bk1}
	b_1=\frac{1}{x};\quad b_{k>1}=\frac{1}{x-b_{k-1}}
	\end{equation}
	
	\noindent Note that in the previous expressions, the index $\kappa$ enumerates the SEs starting from the last one (the $N$-th), while the index $k$ enumerates SEs starting from the zeroth SE. The obtained solutions can be compactly written in a single expression, Eq.(\ref{Vk}):
	
	\begin{equation}\label{Vk}
		\tilde{V}_k=b_{N-k+1}b_{N-k+2}\cdot ...\cdot b_N\tilde{V}_0;\quad k\in\left\{1,2,3,...,N\right\}
	\end{equation}
	
	By repeating this procedure for the LT of the CFs corresponding to the SEs located on the left side of the MC from the zeroth SE, we obtain:
	
	\begin{equation}\label{Uk}
	\tilde{U}_k=c_{M-k+1}c_{M-k+2}...c_M\tilde{V}_0;\quad k\in\left\{1,2,3,...,M\right\}
	\end{equation}
	
	\noindent Here, the coefficients satisfy the rule $c_k=b_k$.
	
	At the end, it is necessary to relate $\tilde{V}_0(x)$  to the initial condition $V_0(\tau=0)$. By combining the expressions $\tilde{V}_1=b_N\tilde{V}_0$, $\tilde{U}_1=c_M\tilde{U}_0$ and equations Eq.(\ref{Vlin0}) i Eq.(\ref{x}),  we easily find the relation between $\tilde{V}_0$ and the initial condition $V_0$:
	
	\begin{equation}\label{V0}
	\tilde{V}_0=i\frac{\hbar\omega_0}{J}\cdot\frac{1}{b_N+c_M-x}V(\tau=0)
	\end{equation}
	
	The dependence of the first few coefficients $b_n(x)$ on the complex variable $x$ is presented in Table \ref{tab:table2b}

	\begin{table}[H] 
		\caption{\label{tab:table2b} Coefficients $b_n(x)$}
		\begin{ruledtabular}
			\begin{tabular}{|l|l|}
				$n$ &$b_n(x)$\\
				\hline\hline
				1 &$b_1(x)=\frac{1}{x}$ \\
				& \\
				2 &$b_2(x)=\frac{x}{x^2-1}$ \\
				& \\
				3 &$b_3(x)=\frac{x^2-1}{x^3-2x}$ \\
				& \\
				4 &$b_4(x)=\frac{x^3-2x}{x^4-3x^2+1}$ \\
				& \\
				5 &$b_5(x)=\frac{x^4-3x^2+1}{x^5-4x^3+3x}$ \\
				& \\
				6 &$b_6(x)=\frac{x^5-4x^3+3x}{x^6-5x^4+6x^2-1}$ \\
				& \\
				7 &$b_7(x)=\frac{x^6-5x^4+6x^2-1}{x^7-6x^5+10x^3-4x}$ \\
				& \\
				8 &$b_8(x)=\frac{x^7-6x^5+10x^3-4x}{x^8-7x^6+15x^4-10x^2+1}$ \\
				& \\
				9 &$b_9(x)=\frac{x^8-7x^6+15x^4-10x^2+1}{x^9-8x^7+21x^5-20x^3+5x}$ \\
				& \\
			\end{tabular}
		\end{ruledtabular}
	\end{table}

	\section{Relation between the coefficients $b_n(x)$ and the modified Chebyshev polynomials of the second kind}
	
	Let us remark that the expressions given in the Table \ref{tab:table2b} can be written in the following form:
	
	\begin{equation}\label{bk2}
	b_n(x)=\frac{D_{n-1}(x)}{D_n(x)}
	\end{equation}
	
	\noindent where $n=1,2,\ldots,N$ and the polynomials $D_n(x)$ appearing in the expressions can be generated by the formula
	
	\begin{equation}\label{BobinaFormula1}
	D_n(x)=\sum_{j=1}^{\left\lfloor\frac{n+2}{2}\right\rfloor}(-1)^{j-1}\binom{n+1-j}{j-1}x^{n+2-2j}
	\end{equation}
	
	\noindent for $n = 0,1,2,\ldots,N$. In Eq.(\ref{BobinaFormula1}), the notation $\left\lfloor \frac{n+2}{2} \right\rfloor$ denotes the greatest integer less than or equal to $\frac{n+2}{2}$. The first few polynomials $D_n(x)$ are listed in Table \ref{tab:table2c}. 
	
		\begin{table}[h]
		\caption{\label{tab:table2c} Polynomials $D_n(x)$}
		\begin{ruledtabular}
			\begin{tabular}{l}
				Polinom $D_n(x)$ \\
				\hline\hline
				\\
				$D_0(x)=1$ \\
				\\
				$D_1(x)=x$ \\
				\\
				$D_2(x)=x^2-1$ \\
				\\
				$D_3(x)=x^3-2x$ \\
				\\
				$D_4(x)=x^4-3x^2+1$ \\
				\\
				$D_5(x)=x^5-4x^3+3x$ \\
				\\
				$D_6(x)=x^6-5x^4+6x^2-1$ \\
				\\
				$D_7(x)=x^7-6x^5+10x^3-4x$ \\
				\\
				$D_8(x)=x^8-7x^6+15x^4-10x^2+1$ \\
				\\
				$D_9(x)=x^9-8x^7+21x^5-20x^3+5x$ \\
				\\
				$D_{10}(x)=x^{10}-9x^8+28x^6-35x^4+15x^2-1$ \\
				\\
				$D_{11}(x)=x^{11}-10x^9+36x^7-56x^5+35x^3-6x$ \\
				\\
			\end{tabular}
		\end{ruledtabular}
	\end{table}
	
	Taking into account Eq.(\ref{bk1}) and Eq.(\ref{bk2}), we easily obtain the following recurrence relation:
	
	\begin{equation}\label{RekDn}
	D_n(x)=xD_{n-1}(x)-D_{n-2}(x);\quad n\in\left\{2,3,\ldots\right\}
	\end{equation}
	
	\noindent which, together with $D_0(x)=1$ and $D_1(x)=x$, generates the remaining polynomials $D_n(x)$. This expression enables straightforward computation of any $n$th--order polynomial with $n >1$, and in this sense, it may serve as an alternative to Eq.(\ref{BobinaFormula1}).
	
	Moreover, for $x\in\mathbb{R}$, by making the substitution $x=2y$, it is easy to see that $U_n(y)=D_n(x)$, where $U_n(y)$ are the Chebyshev polynomials of the second kind \cite{AbramovicStegun}. Therefore, the polynomials $D_n(x)$ can be referred to as \textit{modified Chebyshev polynomials of the second kind}. The roots of the modified Chebyshev polynomials $D_n(x)$ of the second kind are given by:
	
	\begin{equation}\label{DnRoot}
	x_k=2\cos\left(\frac{k}{n+1}\pi\right);\quad k\in\left\{1,2,...,n\right\}
	\end{equation}
	
	
	It is also not difficult to verify that these polynomials satisfy the following recurrence relation:
	
	\begin{widetext}
	\begin{equation}\label{BobinaFormula2}
	D_n(x)D_{m+1}(x)+D_m(x)D_{n+1}(x)-xD_{m+1}(x)D_{n+1}(x)=-D_{m+n+3}(x)
	\end{equation}
	\end{widetext}	
	
	\noindent Some useful properties of modified Chebyshev polynomials relevant to our work are listed here:
	
	\begin{enumerate}
		\item for $n\neq 0$ polynomial $D_n(x)$ has $n$ real roots located within the interval $\left[-2,2\right]$;
		\item if $n$ is even, the roots can be grouped into pairs $(x_i,\; x_{i+1}=-x_i)$;
		\item if $n$ is odd, one root is $x_1 = 0$; the remaining ones can be grouped as in 2);
		\item for large $n$ ($n \gg 1$), the roots accumulate near and tend toward $x = \pm 2$.
	\end{enumerate}

		\nocite{*}
		\bibliography{aipsamp}

\begin{thebibliography}{99}
			
			\bibitem{Voet}
			D. Voet and J. G. Voet, {\it Biochemistry}, (3rd ed.) (Wiley, New York, 2004).
			
			\bibitem{DavydovBQM}
			A. S. Davydov, {\it Biology and Quantum Mechanics} (Naukova Dumka, Kiev, 1979).
			
			\bibitem{Lehninger}
			D. L. Nelson, M. M. Cox, {\it Lehninger Principles of Biochemistry} (W. H. Freeman and Company, New York, 2013).
			
			\bibitem{Petrov}
			E. G. Petrov, {\it Physics of Charge Transfer in Biological Systems} (Naukova Dumka, Kiev, 1984).
			
			\bibitem{Dauxois}
			T. Dauxois and M. Peyrard, {\it Physics of Solitons} (Cambridge University Press, Cambridge, 2006).
			
			\bibitem{Frohlich}
			H. Fr\"{o}hlich and F. Kremer, (Eds.) {\it Coherent Excitations in Biological Systems} (Berlin: Springer, Berlin, 1983).
			
			\bibitem{CruzeiroLTP}
			L. Cruzeiro, Low Temp. Phys. {\bf 48}, 973 (2022).
			
			\bibitem{RSOS}
			D. Chevizovich, D. Micheletto, A. Mvongo, F. Zakiryanov, S. Zdravkovic
			R. Soc. Open Sci. {\bf 7} 200774 (2020).
			
			\bibitem{ChenACIE}
			Xi Chen, Xue Zhang, Xiao Xiao, Zhijia Wang, and Jianzhang Zhao,
			Angew. Chem. Int. Ed. {\bf 62}, e202216010 (2023).
			
			\bibitem{DavydovSMS}
			A. S. Davydov, {\it Solitons in Molecular Systems} (Dordrecht: Reidel, 1985).
			
			\bibitem{PetrovJTB}
			V. N. Kharkyanen, E. G. Petrov and I. I. Ukrainskii, J. Theor. Biol. {\bf 73}, 29 (1973).
			
			\bibitem{BrizhikPRE2014}
			L. S. Brizhik, B. M. A. G. Piette, W. J. Zakrzewski, Phys. Rev. E {\bf 90}, 052915 (2014).
			
			\bibitem{HolsteinAP1}
			T. Holstein, Annals of Physics {\bf 8}, 325 (1959).
			
			\bibitem{HolsteinAP2}Moreover
			T. Holstein, Annals of Physics {\bf 8}, 343 (1959).
			
			\bibitem{ScottPR}
			A. C. Scott, Phys. Rep. {\bf 217}, 1 (1992).
			
			\bibitem{Schuttler}
			H. B. Schuttler, T. Holstein, Ann. Phys. {\bf 166}, 93 (1986).
			
			\bibitem{CastroNetoCaldeira}
			A. H. Castro Neto, A. O. Caldeira, Phys. Rev. B {\bf 46}, 8858 (1992).
			
			\bibitem{ChevizovichPEPAN2025}
			D. Cevizovic, S. Galovic, V. Matic, S. Eh. Shirmovsky, D. V. Shulga, and A. V. Chizhov, Phys. Pat. Nucl. {\bf 56}, 978 (2025).
			
			\bibitem{LF}
			I. G. Lang, Yu. A. Firsov, Zh. Eksp, Teor. Fiz. {\bf 43}, 1843 (1962).
			
			\bibitem{AK}
			D. M. Alexander and J. A. Krumhansl Phys. Rev. B {\bf 33},  7172 (1986).
			
			\bibitem{AlvermanPRB}
			A. Alvermann, H. Fehske and S. A. Trugman, Phys. Rev. B {\bf 81}, 165113 (2010).
			
			\bibitem{YarkonyJCP}
			D. Yarkony, R. Silbey, J. Chem. Phys. {\bf 65}, (1976) 1042.
			
			\bibitem{ZdravkovicCevizovicND}
			D. Chevizovich in: S. Zdravkovic, D. Chevizovich (Eds.), {\it Nonlinear Dynamics of Nanobiophysics} (Springer Nature, Singapore, 2022).  
			
			\bibitem{CevizovicPRE}
			D. Cevizovic, S. Galovic, Z. Ivic, Phys. Rev. E {\bf 84}, 011920 (2011).
			
			\bibitem{FalvoPouthier}
			C. Falvo, V. Pouthier, J. Chem. Phys. {\bf 123},  184709 (2005).
						
			\bibitem{EminPRB}
			D. Emin, Phys. Rev. B \textbf{33}, 3973 (1986).
			
			\bibitem{GogolinPR1988}
			A. A. Gogolin, Phys. Rep. {\bf 157} 347 (1988).
			
			\bibitem{BI}
			D. W. Brown, Z. Ivi\'c , Phys. Rev. B {\bf 40}, 9876 (1989).
			
			\bibitem{HammTsironisEPJST147}
			P. Hamm and G. P. Tsironis, Eur. Phys. J. Spec. Top. {\bf 147} 303 (2007).
			
			\bibitem{HammTsironis}
			P. Hamm, G. P. Tsironis, Phys. Rev. B {\bf 78} 092301 (2008).
						
			\bibitem{IvicCP426}
			D. Cevizovic, Z. Ivic, Z. Przulj, J. Tekic, D. Kapor, Chem. Phys. {\bf 426}, 9 (2013).
			
			\bibitem{ASG}
			A. Szent--Gy{\"o}rgyi, {\it Bioenergetics}, (Academic Press, New York, 1957)
			
			\bibitem{PouthierJCP132}
			V. Pouthier, J. Chem. Phys. {\bf 132}, 035106 (2010)
			
			\bibitem{PouthierPRL}
			J. Edler, R. Pfister, V. Pouthier, C. Falvo and P. Hamm, Phys. Rev. Lett. {\bf 93}, 106405 (2004). 
			
			\bibitem{KalosakasPRB}
			G. Kalosakas, S. Aubry, G. P. Tsironis, Phys. Rev. B {\bf 58}, 3094 (1998).
			
			\bibitem{PouthierPRE2008}
			V. Pouthier, Phys. Rev. E {\bf 78}, 061909--1 (2008).
			
			\bibitem{HennigPRB}
			D. Hennig, Phys. Rev. B {\bf 65}, 174302 (2002).
			
			\bibitem{Rashba}
			E. I. Rashba, in: E. I. Rashba, M. Struge (Eds.), {\it Excitons} (North--Holland, Amsterdam, 1982).
			
			\bibitem{PhysB2005Ivic}
			Z. Ivic, S. Zekovic, D. Cevizovic, D. Kostic, Phys. B {\bf 355}, 417 (2005).
			
			\bibitem{Mahan}
			G. D. Mahan, {\it Many particle Physics}, (Plenum, New York, 1986.)
			
			\bibitem{Nevskaya}
			N. A. Nevskaya and Yu. N. Chirgadze, Biopolymers {\bf 15}, 637 (1976).
			
			\bibitem{KalosakasPRE}
			G. Kalosakas, Phys. Rev. B {\bf 84}, 051905 (2011).
						
			\bibitem{AbramovicStegun}
			M. Abramowitz, I. A. Stegun {\it Handbook of Mathematical Functions with Formulas, Graphs, and Mathematical Tables}, (National Bureau of Standards, 1972)
						
		\end{thebibliography}

	\end{document}